\documentclass[useAMS,usenatbib]{mn2e}

\usepackage[T1]{fontenc}
\usepackage{aecompl}
\usepackage{times}

\usepackage{macros_dc}
\usepackage{astrobib_mnras2e}
\usepackage{graphicx}
\usepackage{epsfig}
\usepackage{amsmath}
\usepackage{hyperref}

\newcommand{\mgroup}{M_{\mathrm{group}}}

\newcommand{\ben}{\begin{enumerate}}
\newcommand{\een}{\end{enumerate}}

\title[Colour-dependent Occupation Statistics]
      {Assessing Colour-dependent Occupation Statistics Inferred from Galaxy Group Catalogues}
\author[Campbell et al.]
{\parbox[t]{\textwidth}{
Duncan Campbell$^{1}$\thanks{E-mail: duncan.campbell@yale.edu}, 
Frank C van den Bosch$^{1}$, 
Andrew Hearin$^{2}$,\\
Nikhil Padmanabhan$^{2}$, 
Andreas Berlind$^{3}$, 
H. J. Mo$^{4}$,
Jeremy Tinker$^{5}$, and \\ 
Xiaohu Yang$^{6,7}$} \\
\vspace*{3pt} \\
$^{1}$Department of Astronomy, Yale University, New Haven, CT 06511, USA \\
$^{2}$Dept. of Physics, Yale University, 260 Whitney Ave, New Haven, CT 06520 \\
$^{3}$Department of Physics and Astronomy, Vanderbilt University, Nashville, TN 37235 \\
$^{4}$Department of Astronomy, University of Massachusetts, Amherst MA 01003-9305, USA \\
$^{5}$Center for Cosmology and Particle Physics, Department of Physics, New York University \\
$^{6}$Center for Astronomy and Astrophysics, Shanghai Jiao Tong University, Shanghai 200240, China \\
$^{7}$Key Laboratory for Research in Galaxies and Cosmology, Shanghai Astronomical Observatory; Nandan Road 80, \\  
\: Shanghai 200030, China
}

\begin{document}

\date{Accepted to MNRAS May 12, 2015}

\pagerange{\pageref{firstpage}--\pageref{lastpage}} \pubyear{2015}

\maketitle

\label{firstpage}

\begin{abstract}
We investigate the ability of current implementations of galaxy group finders to recover colour-dependent halo occupation statistics.  To test the fidelity of group catalogue inferred statistics, we run three different group finders used in the literature over a mock that includes galaxy colours in a realistic manner. Overall, the resulting mock group catalogues are remarkably similar, and most colour-dependent statistics are recovered with reasonable accuracy.  However, it is also clear that certain systematic errors arise as a consequence of correlated errors in group membership determination, central/satellite designation, and halo mass assignment. We introduce a new statistic, the halo transition probability (HTP), which captures the combined impact of all these errors.
As a rule of thumb, errors tend to equalize the properties of distinct galaxy populations (i.e. red vs. blue galaxies or centrals vs. satellites), and to result in inferred occupation statistics that are more accurate for red galaxies than for blue galaxies.  A statistic that is particularly poorly recovered from the group catalogues is the red fraction of central galaxies as function of halo mass. Group finders do a good job in recovering galactic conformity, but also have a tendency to introduce weak conformity when none is present. We conclude that proper inference of colour-dependent statistics from group catalogues is best achieved using forward modelling (i.e., running group finders over
mock data), or by implementing a correction scheme based on the HTP, as long as the latter is not too strongly model-dependent.
\end{abstract}


\section{Introduction} 

The natural site of galaxy formation in the standard $\Lambda$CDM cosmological model of hierarchical structure formation is within gravitationally bound dark matter structures, called dark matter haloes.  Given the hierarchical nature of structure formation, the observed halo substructure present in dark matter N-body simulations, and the long observed propensity of luminous galaxies to live in groups and clusters surrounded by less luminous neighbours, we expect galaxies to group into systems which share a common dark matter halo, a galaxy group.  Galaxy groups suggests a fundamental physical scale important for galaxy evolution, namely the extent of dark matter haloes, and motivate studying galaxies in groups, where group/cluster specific process may influence galaxy evolution, by use of group finders to construct galaxy group catalogues.
  
Spectroscopic redshift surveys are necessary to study galaxies in groups, because precise redshift determinations minimize challenges associated with projection effects on determining galaxy group memberships.  Identifying galaxy groups in redshift surveys has a long history.  Early on, \cite{Turner:1976dw} identified groups in the Catalogue of Galaxies and Clusters of Galaxies followed by \cite{Huchra:1982ed} and \cite{Geller:1983ih} who used the Center for Astrophysics Redshift Survey to search for galaxy groups.  This was followed by other studies identifying galaxy groups in the Las Campanas Redshift Survey \citep{Tucker:2000ig}, the Two Degree Field Galaxy Redshift Survey \citep{Merchan:2002gp, Eke:2004bf,Yang:2005cn}, the Sloan Digital Sky Survey \citep{Merchan:2005gc, Miller:2005kc, Yang:2007db, Berlind:2006dy, Tago:2008kf, Tago:2010jd,  Tinker:2011ek, Tempel:2012er, MunozCuartas:2012cx, Tempel:2014kg} and the Galaxy Mass and Assembly survey \citep{Robotham:2011hu}.

In particular, the Sloan Digital Sky Survey \citep[SDSS,][]{York:2000gn} has been instrumental to the study of galaxy groups by providing the largest precise redshift space galaxy maps to date, along with simultaneous detailed information on galaxy properties. The demographics of galaxies provides a challenge for galaxy evolution models to predict and motivation to understand galaxies' environment.  A colour magnitude diagram of galaxies shows a bi-modality in colour \citep[e.g. see][]{Blanton:2003hh}.  This bi-modality is observed for galaxies, both in the local universe and out to larger redshifts \citep{Bell:2004fo}.  The bi-modality in colour is the result of a bi-modality of specific star formation rates (sSFR), dividing galaxies into a star forming blue cloud and a more quiescent red sequence.  The persistence of this distribution is well established, but the origin of the relation is not well understood.  In addition, many other galaxy properties are correlated without a clear physical explanation, e.g. galaxies with late-type morphologies tend to be star forming, and vice versa for early-type morphologies, and galaxies with increasing stellar mass tend to be quenched.

It is clear that galaxy properties correlate with environment in the local universe.  Galaxies in dense environments such as groups and clusters display an enhanced quenched fraction and an enhanced early-type morphology fraction relative to galaxies in isolated environments \citep{Dressler:1980ie,Postman:1984jx,Balogh:2004cx,Hogg:2004ev,Kauffmann:2004cw,Blanton:2005eb}, and these qualitative trends persist out to higher redshifts \citep{Cucciati:2006fp,Cooper:2007gn,Peng:2010gn}.  It is not clear to what extent a causal relationship exists between environment and galaxy properties and to what extent galaxy properties are determined by processes specific to galaxies in the centre of their own halo. 

Within the context of dark matter halo environments, it has become customary to discuss galaxies in terms of central galaxies and satellite galaxies.  Central galaxies are those galaxies that occupy the centre region of a host halo and are associated with the most massive progenitor of that host halo.  Satellite galaxies are those galaxies that occupy host haloes and are not the central galaxy, generally less massive than the central and associated with dark matter sub-haloes.  It is important to distinguish between these two types of galaxies as they are subject to different physical processes.  A satellite galaxy must have transitioned from being a central in its own host halo to a satellite by crossing into the virial volume of a more massive halo.  The transition from the field prevents dark matter and cold gas from accreting onto the satellite's subhalo \citep{Larson:1980id}.  Satellites falling into this environment can result in the thermalization and/or stripping of gas from the satellite galaxy \citep{Balogh:2000hn, Grebel:2003iw, Kawata:2008cd, McCarthy:2008en}.  Furthermore, satellites are subject to tidal forces from the central potential, stripping mass or disrupting the subhalo, and gravitational interactions with other satellites \citep{Farouki:1981br, Moore:1998jz}.  Meanwhile, central galaxies grow from the cannibalization of satellites, and the accretion of more dark matter and gas \citep{Purcell:2007cn}.  Distinguishing between these two populations allows for the study of the relative importance of these various processes.

From studies utilizing galaxy group catalogues it is clear that quenched fraction and early-type fraction of centrals and satellites increases for increasing host halo mass \citep{Weinmann:2006hu,Yang:2008tg}.  There is also evidence that there is a radial dependence within a halo of satellite quenched fraction \citep{Weinmann:2006hu,Wetzel:2012lk,Wetzel:2014up,Watson:2015gq}, where the quenched fraction falls for increasing radial distance from the halo centre.  It is vital that these trends be determined accurately to provide quantitative constraints on galaxy formation models.  Using constraints from galaxy group catalogues, \citet{Wetzel:2014up} suggest that satellite galaxies quench with a delay time after in-fall into a host halo.  This model predicts a population of ejected satellite galaxies \citep[see also ][]{Teyssier:2012ky} with decreased sSFR, galaxies that have passed through the virial volume of a host halo, but whose orbits have taken them back out beyond the viral radius of the host halo.  The effect of pre-processing \citep{Zabludoff:1998gj}, where satellite galaxies are quenched in a lower mass host halo before being accreted by the current host, can be directly studied with group catalogues.  \citet{Hou:2014yg} use galaxy group catalogues to identify ``sub-groups'', measuring an increased quenched fraction in sub-groups as evidence that pre-processing is important to understand satellite quenching.

Galaxy group catalogues have been used to measure galaxy property correlations beyond those with central/satellite designation and halo mass.  One effect of particular interest here is galactic conformity, first noticed by \cite{Weinmann:2006hu}.  We define such an effect as the tendency of star forming central galaxies to reside in groups with star forming satellites, and vice versa for quenched centrals and satellites, at {\em fixed} halo mass.  Subsequent studies have detected evidence for conformity in the spirit of this original measurement \citep{Kauffmann:2013jd, Phillips:2014bt, Phillips:2014vt, Knobel:2014tx} and at higher redshifts \citep{Hartley:2014uq}.  The robustness of conformity detections and the physical interpretation of conformity is not yet settled.  Adding to this picture, \cite{Kauffmann:2013jd} see further evidence for a correlation of star formation rates between central galaxies and neighbouring galaxies beyond the host halo viral radius, a conceptually different effect than previous measurements of conformity.  We make a distinction between previous measurements, termed 1-halo conformity, and the \cite{Kauffmann:2013jd} large scale phenomenon, termed 2-halo conformity, an effect further elaborated on by \cite*{Hearin:2014wv} as a signature of assembly bias.

The primary goal of a group finding algorithm is to partition a sample of galaxies into the constituent groups.  An ideal group finder would result in a perfect mapping between galaxies which occupy a common halo and a group within a group catalogue.  In reality, because any algorithm is limited to work with observations in redshift space, it is not possible to perfectly assign galaxies to groups.  The result is two classes of errors: first, an algorithm may include a galaxy (or galaxies) which is not part of a halo population into the corresponding group, such galaxies are referred to as interlopers.  Second, an algorithm may misplace a galaxy into an unrelated group.

There has not been a thorough investigation on the accuracy of group catalogues to reproduce unbiased measurements of galaxy properties as a function of halo properties.  \citet{Tinker:2011ek} note that impurities in satellite and central galaxy samples in a group catalogue will bias the quenched fractions measured from group catalogues,and \citet{Hou:2014yg} speculate that group membership contamination may influence the measurements of star formation in galaxy groups as a function of environment.  \cite{Duarte:2014ue} recently developed a method to take into account the uncertainties in the group finding process; however, it is unclear if this will improve the situation.  In this paper, we investigate, in detail, the ability of group finders to recover galaxy(-group) properties as a function halo properties, and how group catalogue failures skew measurements.                    

We choose a set of group finding algorithms from the literature and describe them along with the mock we use to test them in \S~\ref{sec:methodology}.  We then discuss and quantify the errors group finders make in \S~\ref{sec:group_finding_errors}, before we compare the inferred occupation statistics from the group catalogues to the true occupation statistics in \S~\ref{sec:results}.  We conclude with a discussion of our results and a summary in \S~\ref{sec:discussion} and  \S~\ref{sec:summary}.

\section{Methodology}
\label{sec:methodology}

The main goal of this paper is to assess how well group finding algorithms can recover colour-specific occupation statistics of the galaxy population. To that extent, we run three different group finders over a realistic mock galaxy redshift survey, and compare the inferred group statistics to the true, underlying trends in the mocks.  In the following,  we describe the construction of our mock redshift survey (\S\ref{sec:mocks}) and the different group finding algorithms (\S\ref{sec:group_finders}) used in this study.       

\subsection{Mock Catalogues}
\label{sec:mocks}

In order to assess the accuracy with which group finders can recover colour-dependent group statistics of the galaxy population, we need realistic mock catalogues that incorporate galaxy colours.  We construct such a mock by populating dark matter haloes in a large $N$-body simulation (dark matter only) with galaxies of different luminosities and colours, using both subhalo abundance matching \citep{Kravtsov:2004fi, Vale:2004bb, Tasitsiomi:2004gg, Conroy:2006iz, Shankar:2006cd, TrujilloGomez:2011js, RodriguezPuebla:2012ku, Watson:2012jp} and age-matching \citep{Hearin:2013km}, as described below. These techniques have the advantage that, by construction, the mock population has exactly the same luminosity-distribution and colour-distribution as the real data.  In addition, as numerous studies have shown, abundance matching and age-matching are also extremely successful in reproducing various 2-point statistics (e.g., galaxy correlation functions, the excess surface densities inferred from galaxy-galaxy lensing) indicating that the galaxies are placed in the correct dark matter haloes. Furthermore, age-matching `naturally' reproduces galactic conformity, both on small (`1-halo') and large (`2-halo') scales \citep*{Hearin:2014wv}.

In what follows, we present a more detailed description of the numerical simulation used, and the methods used to populate the haloes in the simulation with galaxies.

\subsubsection{Numerical Simulation}
\label{sec:numsim}

The numerical $N$-body simulation used for the construction of our mock catalogues is the Bolshoi simulation \citep{Klypin:2011bd}, which follows the evolution of $2048^3$ dark matter particles using the Adaptive Refinement Tree (ART) code \citep*{Kravtsov:1997iy} in a flat $\Lambda$CDM cosmology with parameters $\Omega_{\rmm,0} = 1 - \Omega_{\Lambda,0} = 0.27$, ${\Omega}_{\rmb,0} = 0.0469$, $n_\rms = 0.95$, $\sigma_8 = 0.82$, and $h = H_0/(100 \kmsmpc) = 0.7$ (hereafter `Bolshoi cosmology'). The box size of the Bolshoi simulation is $L_{\rm box} = 250 \mpch$, resulting in a particle mass of $m_\rmp = 1.35 \times 10^8 \msunh$.

We use the publicly available redshift zero halo catalogue\footnote{\url{http://hipacc.ucsc.edu/Bolshoi/MergerTrees.html}} obtained using the phase-space halo finder ROCKSTAR \citep{Behroozi:2013cn, Behroozi:2012dz}, which uses adaptive, hierarchical refinement of friends-of-friends groups in six phase-space dimensions and one time dimension. As demonstrated in \cite{Knebe:2011jc, Knebe:2013bp}, this results in a very robust tracking of (sub-)haloes \citep*[see also][]{vandenBosch:2014tl}. Haloes in this catalogue are defined to be spherical volumes centred on a local density peak (SO hereafter), such that the average density inside the sphere is $\Delta_{\rm vir} = 360$ times the mean matter density of the simulation box. The radius of each such sphere defines the virial radius $R_{\rm vir}$ of the halo, which is related to the mass of the halo via $M_{\rm vir} = (4/3)\pi R_{\rm vir}^3 \Delta_{\rm vir}\Omega_{m}\rho_{\rm crit}$, where $\rho_{\rm crit} = 3H_0^2/8\pi G$ is the critical energy density of the Universe. Additionally, sub-haloes in this catalogue are distinct, self-bound structures whose centre is found within the virial radius of a more massive host halo. For each host and sub-halo, the catalogue also lists the maximum circular velocity, $V_{\rm max} \equiv {\rm Max}[GM(<r)/r]$, where $M(<r)$ is the mass enclosed within a distance r of the (sub-)halo center, as well as $V_{\rm peak}$, which is defined as the halo's peak value of $V_{\rm max}$ over its entire history.

\subsubsection{Populating Haloes with Galaxies}
\label{sec:populate}

We populate the host haloes and sub-haloes in the Bolshoi simulation with galaxies of Petrosian $r$-band luminosity\footnote{Throughout this paper, all luminosities, magnitudes, and colours are k-corrected to $z=0.1$, using the model described in \cite{Blanton:2003iu}}, $L_r$, using the popular subhalo abundance matching technique, which operates on the premise that there is tight relation between $L_r$ and some property of its dark matter halo. The property considered here is the peak maximum circular velocity, $V_{\rm peak}$. It has been shown that abundance matching works best (i.e. it yields results in closest agreement with observations) for this particular halo parameter \citep{Reddick:2013gi, Hearin:2013ok}.

We start by assigning $r$-band luminosities to haloes and subhaloes using the implicit relation
\begin{equation}
n_{\rm gal}(>L_r) = n_{\rm h}(>V_{\rm peak})\,,
\end{equation}
where $n_{\rm gal}(>L_r)$ is the number density of observed galaxies with $r$-band luminosities larger than $L_r$, which we compute using the SDSS $r$-band luminosity function of \cite{Blanton:2005fs}, and $n_\rmh(>V_{\rm peak})$ is the number density of dark matter haloes and subhaloes with peak circular velocity larger than $V_{\rm peak}$, which we obtain from the Bolshoi simulation. Because of the finite mass resolution of the simulation, and the magnitude limit of the SDSS data, we only assign galaxies to haloes that have $r$-band magnitudes $M_r -5\log(h) < -19$. This results in populating all haloes down to $V_{\rm peak} \sim 100 \kms$.

\begin{figure}
\includegraphics[width=\columnwidth]{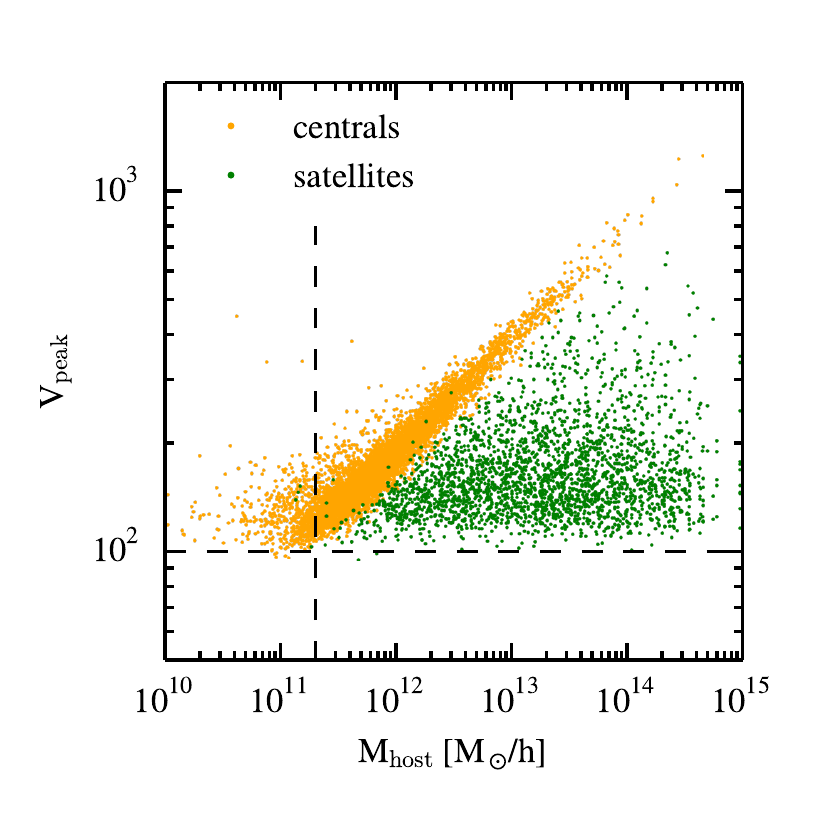}
\caption{A random sub-sampling of host halo mass, $M_{\rm host}$, vs. peak halo circular velocity, $V_{\rm peak}$, of mock galaxies.  The horizontal dashed black line is the $V_{\rm peak}$ cut corresponding to $M_{\rm r} -5\log(h) < -19$.  The vertical dashed black line is the cut on halo mass, $M_{\rm host}$, that results in the equivalent number density of haloes.}
\label{fig:vpeak_mhost}
\end{figure}

After this first step, we add stochasticity to the monotonic relationship between $L_r$ and $V_{\rm peak}$, using the method described in Appendix A of \cite{Hearin:2013km}. Some amount of scatter is expected, given the complex interplay of physical processes in galaxy formation, and results in mocks that are in better agreement with observational data \citep[e.g.,][and references therein]{Klypin:2011bd, TrujilloGomez:2011js, Watson:2012jp} than mocks without scatter. Our model for the stochasticity in the luminosity of mock galaxies results in a uniform scatter in luminosity of $\sim 0.15$ dex at fixed $V_{\rm peak}$. Due to the scatter between $M_{\rm vir}$ and $V_{\rm peak}$, this translates into $\sim0.18$ dex of scatter in luminosity at fixed $M_{\rm vir}$, which is in excellent agreement with observational constraints \citep[e.g.,][]{Cooray:2006ek, Yang:2009cm, More:2011il, Cacciato:2013dd}.  The result of this process is the distribution of $V_{\rm peak}-M_{\rm host}$ for central and satellite galaxies shown in Fig. \ref{fig:vpeak_mhost}.

Finally, we assign $(g-r)$ colours to all galaxies in our mock catalogue using the age-matching technique introduced by \citet{Hearin:2013km} and \citet{Hearin:2014hh}. Age-matching operates on the premise that, at fixed luminosity, there is a monotonic relation between galaxy colour and some proxy for (sub-)halo age.  First, all mock galaxies in a narrow bin of $r$-band luminosity are rank-ordered according to this halo age proxy. Next, for each galaxy in the bin, a colour is drawn from the observed distribution in the SDSS, $P(g-r|L_r)$.  These colours are also rank-ordered, and subsequently each galaxy in the bin is assigned a colour by matching rank-orders, such that the reddest (bluest) colour is assigned to the galaxy whose (sub-)halo has the oldest (youngest) age\footnote{The resulting mock catalogue is publicly available at \url{http://logrus.uchicago.edu/~aphearin}}. We refer the reader to the original papers for details regarding the exact definition of halo age used, but roughly it corresponds to the time when the main progenitor of the (sub-)halo first reaches a mass equal to four percent of its present day mass.  We note that there is no scatter included in the relation between halo age and assigned colour, and we leave an examination of a tunable assembly bias mock to a forthcoming paper.    

\begin{figure}
\includegraphics[width=\columnwidth]{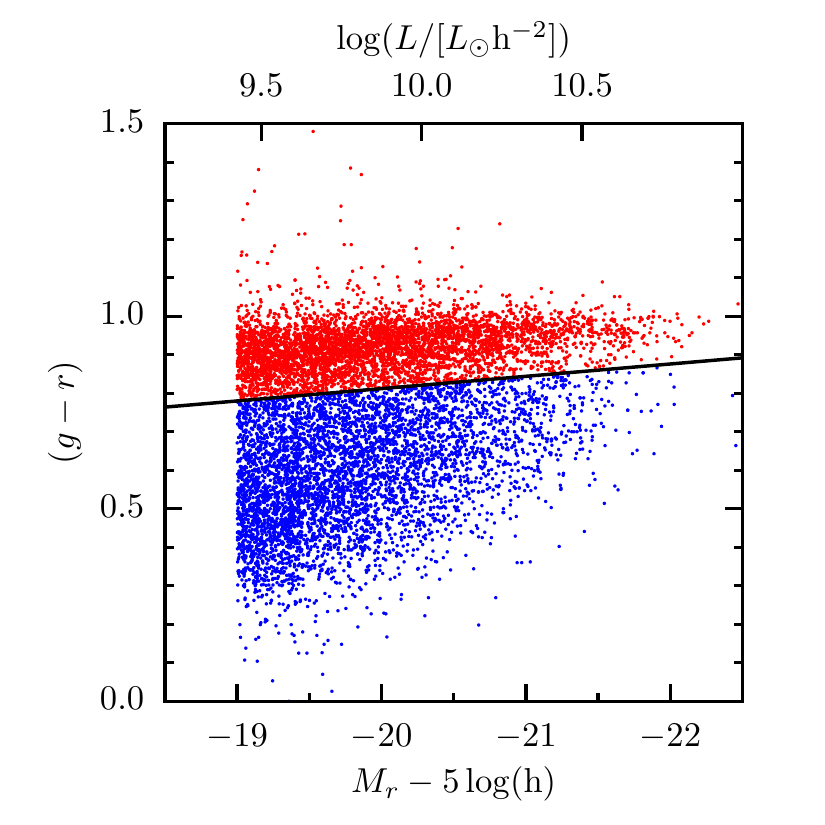}
\caption{A random sub-sampling of the $(g-r)$ colours vs. r-band absolute magnitude of mock galaxies.  The black line [Eq.~\ref{eq:color_cut}] is the colour cut used to define the red (red points) and blue (blue points) sub-samples of galaxies.}
\label{fig1}
\end{figure}

Once all mock galaxies have been assigned luminosities and colours, we split the sample into `red' and `blue' sub-samples, using a magnitude-dependent cut of \citet{Weinmann:2006hu}, which roughly follows the observed bi-modality in the colour-magnitude relation:
\begin{equation} \label{eq:color_cut}
(g-r)_{\rm cut} = 0.7 - 0.032[M_r - 5\log h + 16.5]\,,
\end{equation}
(see Fig.~\ref{fig1}).  In what follows, we refer to galaxies that are redder and bluer than $(g-r)_{\rm cut}$ as `red' and `blue' galaxies, respectively.  We refer the reader to \cite{Taylor:2015im} for a thorough discussion of definitions of `red' and `blue' galaxies, and the consequences of our choice.   

\subsubsection{Mock Redshift Survey}
\label{subsec:redshift_survey}

As the final step, we construct volume limited galaxy redshift surveys from the age-matching mock. We place a virtual observer at one of the corners of the simulation box, define a $(\alpha,\delta)$ (right ascension, declination) coordinate frame, and compute, for each mock galaxy, its angular coordinates, its redshift, $z$ (accounting for the peculiar velocity along the line-of-sight towards the virtual observer), and its apparent magnitudes in the $r$ and $g$ bands, $m_r$ and $m_g$, respectively. Next, we remove all galaxies with $m_r > 17.77$, mimicking the apparent magnitude limit of the SDSS spectroscopic survey.  From the resulting catalogue we construct a volume limited sample for galaxies $M_r-5\log h < -19.0$ for which the following information is available: $(\alpha, \delta, z, m_r, m_g)$.  

We do not incorporate observational errors in these observables, nor do we model the spectroscopic incompleteness of a realistic survey.  \cite{Yang:2007db} find that spectroscopic incompleteness in SDSS has a minimal affect on their results, where it is straight-forward to empirically examine the effect.  We leave a further discussion of this issue to \S~\ref{subsec:primary_prop}.  The magnitudes are simply redshifted Petrosian magnitudes from our mock.  While we are consistent with our use of a magnitude system, it should be noted that the use of a different magnitude system may introduce systematic changes in our mock and therefore in our results for the group finders run on our mock.  We have tested that errors in magnitudes and redshifts similar to those in SDSS have a negligible affect on all of our results.    

\subsection{Group Finding Algorithms}
\label{sec:group_finders}

In order to gauge the accuracy with which group finders can extract colour-dependent statistics from galaxy redshift surveys, we run three different group finders over the mock galaxy catalogue described above. The three group finders used are the friends-of-friends (FoF) based group finder of \citet{Berlind:2006dy}, the halo-based group finder developed by \cite{Yang:2005cn, Yang:2007db}, and the modified version thereof used by \cite{Tinker:2011ek}. The Yang \etal and Tinker \etal group finders assume that dark matter haloes are SO, and use the galaxy luminosities in assigning group memberships. They mainly differ in the starting points used to build the groups.  The Berlind \etal group finder, on the other hand, makes no such assumption about halo shape, and ignores all information regarding galaxy luminosities when partitioning galaxies among groups.  These three group finders are representative of group finders in general; most group finders used in the literature use algorithms (or combinations thereof) that are similar to those employed in the group finders used here.

In the following sections we give a brief description of each group finder's method to assign membership to groups, and we point the reader to the relevant papers for further details on the methods used.  Finally, we describe the specific algorithms used for all three group finders to estimate halo mass and central/satellite designation for individual groups. 

\subsubsection{Berlind et al. group finder}
\label{sec:Berlind}

\citet{Berlind:2006dy} adopt a simple FoF algorithm to identify galaxy groups, which is the most common algorithm used to select groups from a redshift survey \citep[e.g.,][]{Huchra:1982ed, Geller:1983ih, Nolthenius:1987vf, Ramella:1989fz, Ramella:1997kj, Ramella:1999ul, Ramella:2002id, Moore:1993kv, Tucker:2000ig, Giuricin:2000eg, Merchan:2002gp, Eke:2004bf}. Following \cite*{Huchra:1982ed}, a pair of galaxies is linked if both their transverse and line-of sight separations are smaller than a specified pair of projected and line-of-sight linking lengths, respectively. Formally, two galaxies, $i$ and $j$, with an angular separation, $\theta_{ij}$, and observed redshifts, $z_i$ and $z_j$, have a line-of-sight separation given by,
\begin{equation}
D_{\parallel,ij}=\frac{c}{H_0}|z_i-z_j|
\end{equation}  
and a projected separation given by,
\begin{equation}
D_{\perp,ij}=\frac{c}{H_0}(z_i+z_j)\sin{\left(\frac{\theta_{ij}}{2}\right)}.
\end{equation}  
The linking condition is,
\begin{equation}
{D_{\parallel,ij}\leq b_{\parallel}\bar{n}_g^{-1/3}} 
\end{equation}
and
\begin{equation}
{D_{\perp,ij}\leq b_{\perp}\bar{n}_g^{-1/3}},
\end{equation}
where $\bar{n}_g$ is the global, mean number density of galaxies and $b_{\perp}$ and $b_{\parallel}$ are the projected and line-of-sight linking lengths, respectively, in units of the mean inter-galaxy separation. The FoF algorithm is recursive, and links all galaxies that obey the above linking condition to each other, thus yielding a unique group of galaxies. Note that this group finding algorithm uses only galaxy angular positions and observed redshifts, and not galaxy luminosities.

The linking length must be tuned to minimize interlopers and maximize completeness of group members. Typically, linking lengths that are too small will decrease completeness, while linking lengths that are too large will increase the number of interlopers in groups.  \citet{Berlind:2006dy} used mock catalogues to tune their linking lengths, using a number of criteria to judge the quality of the resulting group catalogues. They end up using $b_{\parallel}=0.75$ and $b_{\perp}=0.14$, corresponding to $\sim 3~h^{-1}{\rm Mpc}$ and $\sim 0.6~h^{-1}{\rm Mpc}$ for our sample respectively. We adopt these values for which they find that groups with $N_\rmg \geq 10$ have:
\begin{enumerate}
  \item an unbiased group multiplicity function.
  \item a spurious group fraction less than $\sim 1\%$, where spurious groups are fractured `pieces' with no true central galaxy.
  \item a halo completeness greater than $\sim 97\%$, which implies that fewer than $\sim 3\%$ of the groups have more than one true central among their members
  \item an unbiased projected size distribution as a function of group multiplicity
  \item a velocity dispersion distribution that is $\sim 20\%$ too low at all group multiplicities.
\end{enumerate}  
Note that the performance is expected to be inferior for groups with fewer assigned members. Finally, we stress that this group finder was tuned using mocks in which the dark matter haloes are identified using the FoF algorithm with a linking length of $b=0.2$. In addition, satellite galaxies were assigned the positions and velocities of randomly selected dark matter particles within those FoF haloes, and therefore do not occupy spherical volumes. This is different from the mocks that we use in this paper, which may have implications for our assessment of the performance of the Berlind \etal group finder on our mocks (see discussion in \S~\ref{subsec:discussion_halodef}).

\subsubsection{Yang et al. group finder}
\label{sec:Yang}

\cite{Yang:2005cn, Yang:2007db} developed a halo-based group finder, which has the advantage that it is iterative and based on an adaptive filter modelled after the expected phase-space properties of dark matter haloes.  First, potential group centres and initial group membership estimates are identified using a FoF algorithm with very small linking lengths of $b_{\parallel}=0.3$ and $b_{\perp}=0.05$. The geometrical, luminosity-weighted centers of all the FOF groups thus identified with two members or more are considered as the centres of potential groups. All galaxies not linked to any of these FOF groups are also treated as tentative centers of potential groups. Next, the characteristic group luminosity, $L_{19.5}$, is computed, where $L_{19.5}$ is the total luminosity of all group members with magnitude $M_{\rm r} - 5 \log h \leq -19.5$.  The characteristic group luminosity is used together with an estimate of the group's mass-to-light ratio, $M_{180}/L_{19.5}$, to obtain an initial estimate of the group's halo mass, $M_{180}$.  Note that here haloes are defined as SO with an average overdensity of 180 times the mean background density. In the first iteration, it is simply assumed that {\it all} groups have $M_{180}/L_{19.5} = 500\,h\,M_{\odot}/L_{\odot}$.  For all subsequent iterations, however, the $M_{180}/L_{19.5}$ relation that derives from the previous iteration is used, as described below. As demonstrated in \cite{Yang:2005cn}, this iterative technique makes the final group catalogue very insensitive to this (arbitrary) initial guess for $M_{180}/L_{19.5}$.

Assuming that dark matter haloes follow an NFW density profile \citep*{Navarro:1997if}, and that the distribution of galaxies in phase space follows that of the dark matter particles, the number density contrast of galaxies in the redshift space around the group centre (assumed to coincide with the centre of the halo) can be written as
\begin{equation}
P_M(R,\Delta z) = {H_0 \over c} \, {\Sigma(R) \over \bar{\rho}} 
\, p(\Delta z)\,.
\end{equation}
Here $c$ is the speed of light, $\Delta z = z -z_{\rm group}$, $\bar{\rho}$ is the average density of universe, $\Sigma(R)$ is the projected surface density of a (spherical) NFW profile, and $p(\Delta z) \rmd\Delta z$ describes the redshift distribution of galaxies within the halo and is assumed to have a Gaussian form with a velocity dispersion equal to $\sigma_{180} (1+z_{\rm group})$.  Here $\sigma_{180}$ is the one-dimensional velocity dispersion of an isotropic NFW halo of mass $M_{180}$ (see van den Bosch \etal 2004).

The method re-evaluates group membership by assigning each galaxy to any group for which $P_M(R,\Delta z) \geq B$, where $B$ is a free parameter. If a galaxy can be assigned to more than one group, it is assigned to the group which maximizes $P_M(R,\Delta z)$, and if all the members of one group can be assigned to another, the groups are merged.  Next the group centres are recomputed, and the entire process is iterated until there is no further change in group memberships. Finally, each group is assigned a new halo mass, $M_{180}$, using abundance matching on $L_{19.5}$, and using the \cite{Tinker:2008ja} halo mass function for the Bolshoi cosmology and for the halo definition used here. These newly derived masses are used to update the relation between $M_{180}/L_{19.5}$ and $L_{19.5}$, and the entire procedure is iterated until the $M_{180}$-$L_{19.5}$ relation has converged, which typically takes only three to four iterations.

Using mock galaxy redshift surveys, \cite{Yang:2005cn} tuned the free parameter, $B$, by maximizing a measure of completeness of groups, while minimizing the number of interlopers in groups.  The best-fit results were obtained for $B=10$ for an SDSS main sample like survey, which is also the value used for this study.  In general this will depend on spectroscopic completeness and red-shift error.  We refer the reader to \cite{Yang:2005cn} for a discussion of these effects.  

\subsubsection{Tinker et al. group finder}
\label{sec:Tinker}

The galaxy group finder developed by \cite{Tinker:2011ek} is a modified version of that of \cite{Yang:2005cn} described above.  It starts by estimating initial (sub-)halo masses for each galaxy using the sub-halo abundance matching method.  This method relates the galaxy $r$-band luminosity to the mass of a (sub-)halo by assuming a monotonic relation with the brightest galaxies living in the most massive (sub-)haloes using the halo mass function from \cite{Tinker:2008ja} and the sub-halo mass function from \citet{Tinker:2010ha}.  With this initial (sub-)halo mass estimate for each galaxy, the viral radius and velocity dispersion are determined using:
\begin{equation}
R = \left(\frac{3M_{200}}{4\pi200\bar{\rho}}\right)^{1/3},
\end{equation}
and
\begin{equation}
\sigma_v=\sqrt{\frac{GM_{200}}{2R_{200}}(1+z)},
\end{equation}
where $M_{200}$ is defined as the mass of a SO halo with an average overdensity of $200$ times the mean background density, and $R_{200}$ is the corresponding radius of such a halo.

Once halo properties have been assigned to each galaxy, the probability that the galaxy is a central or satellite galaxy in each group is determined in the same way as the Yang et al group finder, namely that a galaxy is assigned to be a satellite of a group when $P_M(R,\Delta z) \geq B$.   This algorithm is applied to the galaxies with initial halo properties determined using abundance matching.  After the initial group memberships are assigned, halo mass is calculated by abundance matching on total group luminosity and host haloes only.  This procedure is iterated until group memberships remain unchanged.  Group membership is sensitive to the choice of the constant $B$, and this parameter must be chosen to best recover the desired groups, $B=10$.  We do not retune this parameter and note that the group finder was tuned on a mock with haloes defined as SO with a mean internal density of $200$ times the mean density of the universe.

\subsubsection{halo mass and central/satellite designation}
\label{subsec:mass_and_central_assignment}

Once the group finders have been used to assign group memberships, halo mass estimates and central/satellite designations are made as follows.  We assign each group a halo mass, $M_{\rm group}$, using abundance matching on total group luminosity, $L_{\rm group}$, using the \cite{Tinker:2008ja} halo mass function for the Bolshoi cosmology. Note that $L_{\rm group}$ is defined as the summed luminosity of all assigned group members, which, by virtue of the volume limited nature of the mock galaxy redshift survey, are all brighter than $M_r - 5\log h = -19$. The particular mass function used is the one for which haloes are defined as SO with an average internal density 200 times the mean background density, $M_{200}$.  We use zero scatter in the $L_{\rm group}$-$M_{\rm group}$ relation, rank ordering groups by $L_{\rm group}$ and calculating the cumulative number density to assign halo mass estimates such that $n_{\rm group}(>L_{\rm group})=n_{\rm halo}(>M_{200})$.  Finally, a galaxy is designated as a central galaxy if it is the brightest group member, otherwise it is designated as a satellite.

While we assign consistent halo mass estimates to groups in all three group finders using the algorithm described above, each group finder used in this study uses a different halo definition when tuning each respective algorithm.  We briefly summarize these differences as follows and leave a discussion of the consequences to a discussion in \S~\ref{subsec:discussion_halodef}.  The Berlind et al. group finder is tuned and tested on a mock where haloes are defined using an FoF algorithm with a linking length 0.2 times the mean interparticle separation.  The Yang et al. group finder is tuned on a mock where haloes are defined as SO with a mean internal density of 180 times the background density, while the Tinker et al. group finder is tuned on a mock where haloes are defined as SO with a mean internal density of 200 times the background density.  None of these definitions is the same as the one used to {\em populate} galaxies in our mock, where haloes are defined as SO with a mean internal density of 360 times the background density.  We do not retune any group finder for this study, and instead use each group finder {\em as is}.  To facilitate comparison, we assign halo mass estimates to groups with a single algorithm, and use the same halo definition, namely $M_{200}$, as described in the previous paragraph. When we make comparisons between group finder results and the mock, we use a $M_{200}$ halo mass measurement for haloes in the mock for consistency.

Finally, our mock and group finders assume that galaxy luminosity is the primary galaxy property which drives the galaxy occupation statistics.  This is a common assumption, but stellar mass could instead have been assumed to be the primary property.  In line with our mock, for our analysis we use galaxy luminosity to estimate halo mass as described previously.  Our choice minimizes systematic errors associated with uncertainty in what property is most applicable to the real universe, and thus our results represent a best-case scenario for a group catalogue analysis.  In general, differing assumptions between the mock and grouping analysis gives rise to systematic errors.  We leave a discussion of this to \S~\ref{subsec:primary_prop}, but we note that the systematic uncertainty in statistics dependent on halo mass can be substantially impacted by this issue.

\section{Group Finder Errors}
\label{sec:group_finding_errors}

In the following section we describe the manner in which group finding algorithms can fail (\S~\ref{subsec:generic_failures}), the classic purity and completeness metrics (\S~\ref{subsec:purity_completeness}), and finally we introduce a new technique to comprehensively characterize group-finder errors: the halo transition probability (HTP, \S~\ref{subsec:HTP}).  When discussing the errors made by group finders in partitioning galaxies into groups, it is often necessary to make a distinction between the groups identified by a group finder and the {\em true} groups present in the mock.  When such a distinction could be ambiguous, we reserve the term `group' or `g' for the result of a group finder and use the term `halo' or `h' to refer to the truth.  Furthermore we will refer to the three group finders in shorthand for the remainder of the paper as {\tt BERLIND-GF}, {\tt YANG-GF}, and {\tt TINKER-GF} for the Berlind et al., Yang et al., and Tinker et al. group finders respectively.

\subsection{Generic Failure Modes of Group Finders}
\label{subsec:generic_failures}

\subsubsection{membership allocation errors}

\begin{figure}
\includegraphics[]{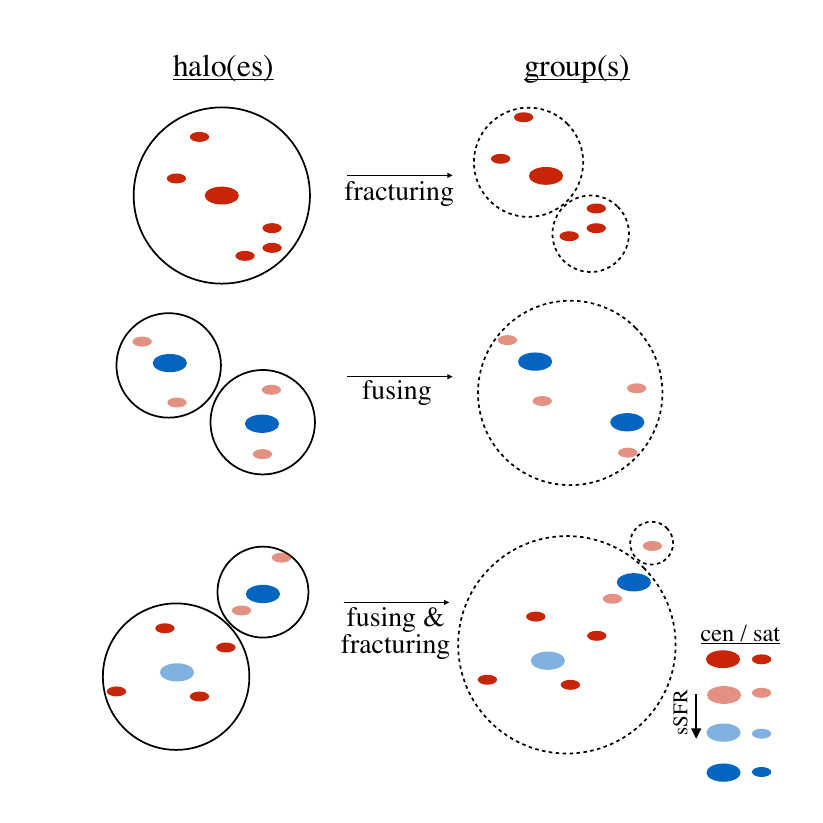} 
\caption{Illustration of group finder failure modes: fracturing (top), fusing (middle), and both simultaneously (bottom).  Solid circles on the left denote the boundaries of haloes and the coloured points within the circle indicate galaxies that truly live in that halo.  Dashed circles on the right indicate the boundaries of group finder identified haloes and the coloured points within the dashed circles comprise galaxy groups.  The size of the bounding circle may be interpreted as halo/group mass, the colour of filled circles as sSFR, and the size as an indication of whether it is a central/satellite galaxy in its host halo.}
\label{fig:artwork}
\end{figure}

The first challenge of any group finder is to correctly partition galaxies into their groups (i.e. to correctly identify group memberships). We take a similar approach as \cite{Duarte:2014ug} and identify two failure modes associated with this step, which we call `fracturing' and `fusing'\footnote{\cite{Duarte:2014ug} use the terms `fragmentation' and `merging' instead of `fracturing' and `fusing', respectively.} (see Fig. \ref{fig:artwork}).  A group has been fractured if galaxy members of a common halo have been placed in two or more distinct groups.  A group has been fused with another group if any members from two distinct haloes have been assigned to one and the same group.  Note that a single group/halo can experience both fracturing and fusing simultaneously, as illustrated in Fig.~\ref{fig:artwork}.  The effect of fusing and fracturing of a group will be key to explaining the effects of group finding errors.  From the outset, one should expect fusing to result in a galaxy from a low mass halo being identified in a higher mass group and vice versa for fracturing.

\subsubsection{central/satellite designation errors}
\label{subsec:cen_sat_errors}

The second challenge for a group finder is to identify the central galaxy of a group. In all three group finders considered here, we follow \cite{Yang:2007db} and \cite{Tinker:2011ek} and identify the central galaxy as the brightest group member. Any group member that is not a central galaxy is subsequently identified as a satellite. Hence, every group will have one, and only one, central galaxy, while the number of satellite galaxies can be any positive integer, including zero.  

This method for selecting centrals and satellites is motivated by the idea that central galaxies grow in mass by cannibalizing their satellites \citep{Dubinski:1998bz, Cooray:2005gd} and by being the repositories of cooling flows, whereas satellite galaxies are subjected to a number of processes that quench star formation (e.g., ram-pressure stripping, strangulation) and strip mass (e.g. tidal stripping). If this methodology is adopted in galaxy group catalogues, the resulting populations of centrals and satellites are clearly distinct, in that they have different properties (e.g., colours, star formation rates, AGN activity, morphologies) at fixed luminosity or stellar mass \citep[e.g.,][]{Weinmann:2006hu, Skibba:2007fh, vonderLinden:2007ev, vandenBosch:2008fv, Pasquali:2009fi, Pasquali:2010ky, Skibba:2009kv, Hansen:2009if}. In addition, It has been shown that brightest group members do not obey extreme value statistics, indicating that they are truly a special class among the entire population of galaxies \citep[see][and references therein]{Hearin:2013ok, Shen:2014kq}.  Although this indicates that identifying the brightest group members as centrals correctly identifies centrals from satellites {\it in a statistical sense}, it is unlikely to be correct in each and every group. Indeed, \cite{vandenBosch:2005jo} and \cite{Skibba:2011dx}, using the phase-space statistics of brightest group galaxies, have shown that in a significant fraction of dark matter haloes, ranging from $\sim 25$ percent for Milky Way size haloes to $\sim 40$ percent for cluster size haloes, the brightest group member is a satellite rather than the true central.  In haloes where a satellite is truly brighter than the central, our method to identify centrals is guaranteed to fail.  

\begin{figure}
\includegraphics[width=\columnwidth]{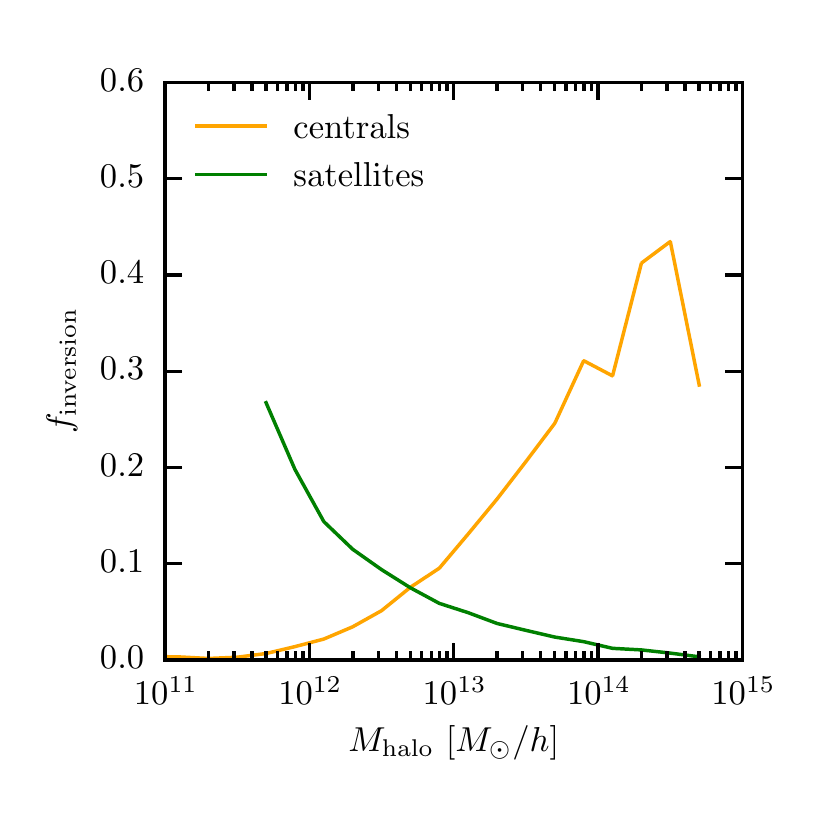}
\caption{orange: the central inversion fraction for groups as a function of halo mass, i.e. the fraction of groups whose brightest member is not a central galaxy.  green: the satellite inversion fraction for groups, i.e. the fraction of satellites that are the brightest member in a group.  For groups where the true central is not the most luminous member, our criteria to identify centrals will result in an error. }
\label{fig:inversion_fraction}
\end{figure}

To characterize this phenomenon in our mock, we define the central inversion fraction, the fraction of groups whose brightest galaxy is not a central galaxy in the mock.  In the limit where group membership is determined perfectly, the central inversion fraction can be written as:
\begin{equation}
f_{\rm cen, inv}=\frac{N_{\rm cen|sat}}{N_{\rm groups}},
\end{equation}   
where $N_{\rm cen|sat}$ is the number of true satellites identified as centrals.  A closely related quantity is the satellite inversion fraction:
\begin{equation}
f_{\rm sat, inv}=\frac{N_{\rm sat|cen}}{N_{\rm sat}},
\end{equation}   
From the intrinsic $f_{\rm inversion}$ in the mock (see Fig. \ref{fig:inversion_fraction}) we can read off the minimum error expected in identifying centrals/satellites given the designation criterion chosen for this study.  If groups were identified perfectly, in groups with a halo mass of $10^{13}~M_{\odot}$, roughly 10\% would have a satellite misidentified as a central.

Note that any fusing or fracturing in the group finding process (as defined above) will generally lead to an even larger number of misclassifications.  The fusing process results in two or more groups being associated with one halo (see middle row in Fig. \ref{fig:artwork}), while fracturing results in one halo being associated with two or more groups (see top row in Fig. \ref{fig:artwork}).  Each of these cases, respectively, will result in at least one central being identified as a satellite, and one satellite being identified as a central.  One of the goals of this paper is to gauge how the misclassification of centrals and satellites in galaxy group catalogues impacts the inferred colour-dependent statistics of each population, e.g. central red fraction and galactic conformity.

\subsubsection{halo mass estimation errors}
\label{subsec:mass_errors}

\begin{figure}
    \includegraphics[width=\columnwidth]{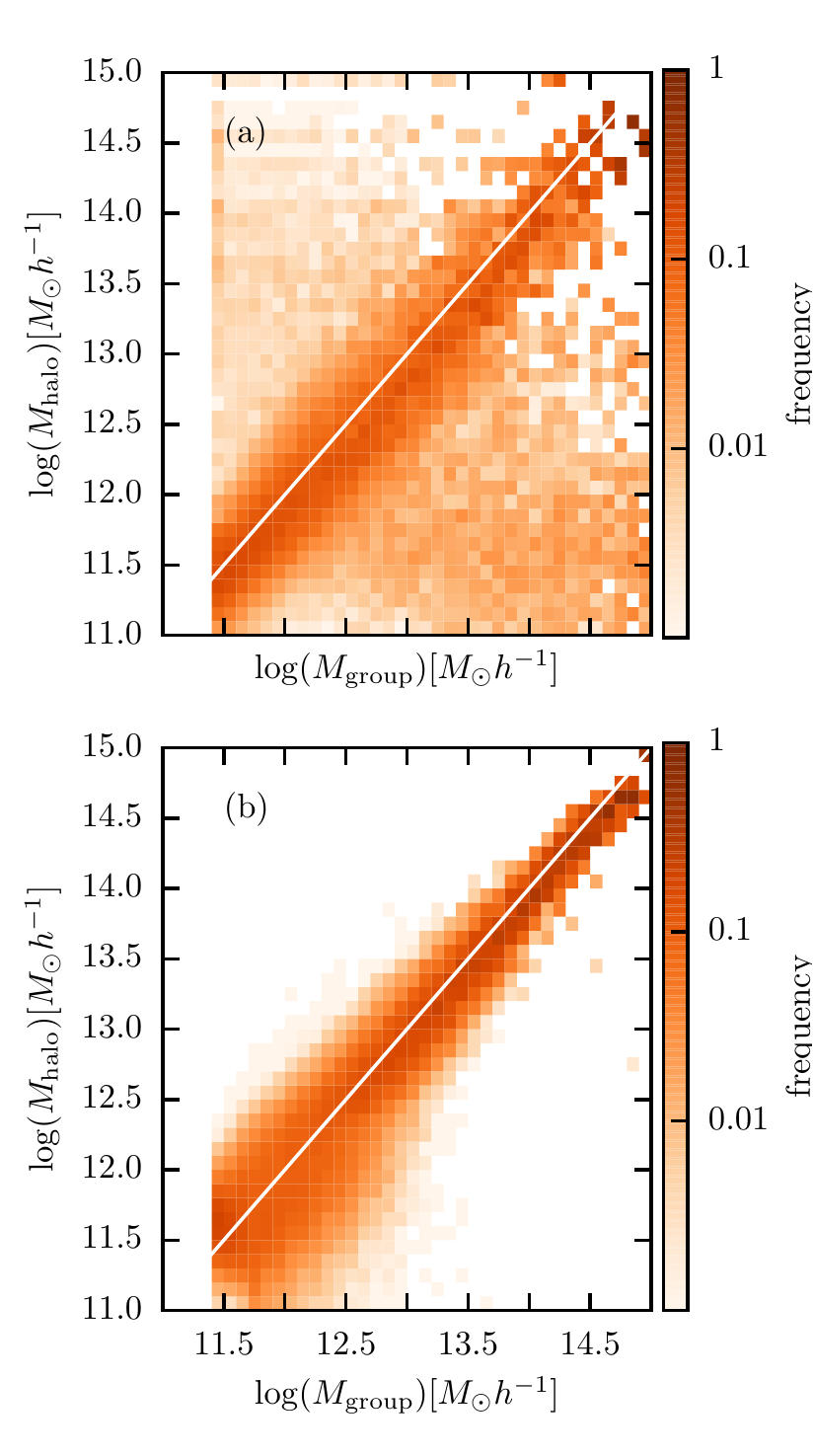}
    \caption{$P(M_{\rm halo}|M_{\rm group})$: the distribution of halo masses, $M_{\rm halo}$, that galaxies assigned to a group of mass, $M_{\rm group}$, are drawn from for the {\tt TINKER-GF} (panel a) and in the case of perfect group membership (panel b), where group luminosity is used to estimate halo mass.}
    \label{fig:mass_HTP}
\end{figure}

The third and final challenge for galaxy group finders is to estimate
the halo mass for each individual group. In principle, this can be
done using a variety of techniques, such as satellite kinematics,
gravitational lensing, or X-ray observations.  In practice, however,
each of these methods is only feasible for the most massive groups and
clusters. Typically, the only available information is group richness
(multiplicity), and the luminosities and line-of-sight velocities of
the member galaxies. In this paper we assign halo masses to each
identified group by using abundance matching on total group luminosity
(see \S~\ref{subsec:mass_and_central_assignment} for details), and we
use exactly the same method for each of the three group finding
algorithms used in this paper.  

\begin{figure*}
  \includegraphics[]{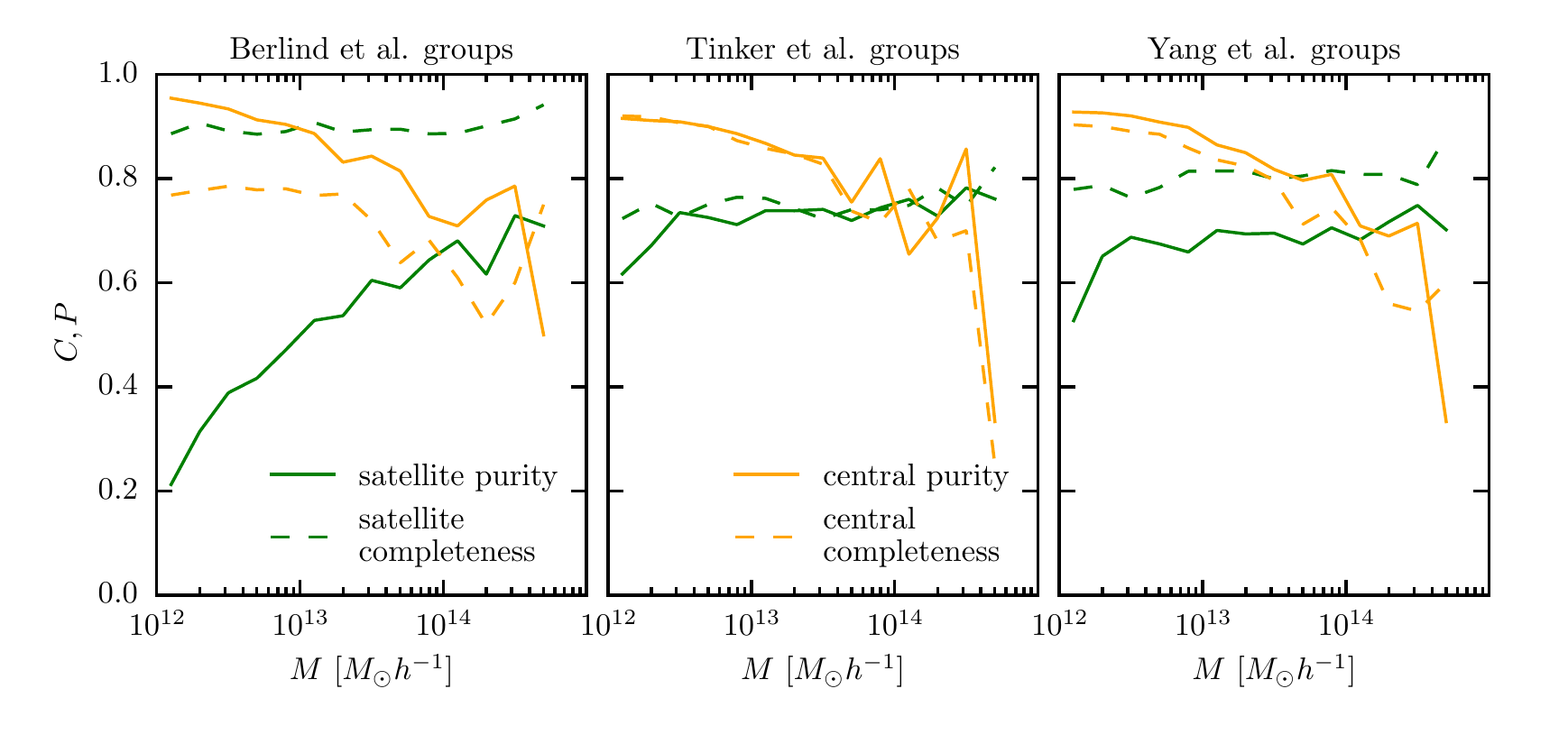}
  \caption{The purity (solid lines) and completeness (dashed line) of centrals (orange) and satellites (green) as a function of group/halo mass for each group finder: {\tt BERLIND-GF} groups (left),  {\tt TINKER-GF} groups (middle), {\tt YANG-GF} groups (right).}
  \label{fig:purity_completeness}
\end{figure*}

This method to estimate halo mass for groups is fundamentally limited
in two ways. First of all, one expects intrinsic scatter in the
relation between group luminosity, $L_{\rm group}$, and halo mass,
$M_{\rm halo}$, which is not accounted for. Secondly, any errors in
group membership determination will generally result in errors in
$L_{\rm group}$, and thus in the inferred halo mass. Typically, a
group that is fractured will have a $L_{\rm group}$ that is too low,
resulting in an underestimate of its halo mass, while the opposite
applies to fused groups. The upper panel of Fig.~\ref{fig:mass_HTP}
shows the relation between true halo mass in the mock, $M_{\rm halo}$,
and the inferred group mass, $M_{\rm group}$, for individual galaxies
in the {\tt TINKER-GF} groups. The results for the other two group
finders are very similar. Note how most galaxies are assigned masses
that are approximately correct, in that they lie close to the 1:1
locus (white line).  However, there is also a population of galaxies
(in the upper left and bottom right regions) for which the estimated
mass is catastrophically wrong. These are galaxies whose groups have
experienced fusing or fracturing. This is evident from the lower panel
of Fig.~\ref{fig:mass_HTP}, which shows the same results, but in the
absence of group membership errors, and which therefore reveals the
scatter in inferred group mass that arises purely from the intrinsic
scatter in the $M_{\rm halo}-L_{\rm group}$ relation.

In general, the error that arises from the intrinsic scatter in the
$M_{\rm halo}-L_{\rm group}$ relation will not have a significant
impact on the {\it average} relation between some group property, $Q$,
and group mass; i.e., the inferred $\langle Q|M_{\rm group} \rangle$
will be similar to the true $\langle Q|M_{\rm halo} \rangle$, {\it as
  long as the scatter in the $M_{\rm halo}-L_{\rm group}$ relation is
  uncorrelated with $Q$}. If, on the other hand, such a correlation
{\it is} present, the intrinsic scatter in the $M_{\rm halo}-L_{\rm
  group}$ relation will introduce a systematic error in the inferred
$\langle Q|M_{\rm group} \rangle$. To see this, consider an example in
which $Q$ is positively correlated with $L_{\rm group}$ at fixed
$M_{\rm halo}$. At that fixed halo mass, groups with a high value of
$Q$ will have a relatively large $L_{\rm group}$, and their inferred
group mass, $M_{\rm group}$, will therefore be systematically
overestimated. Similarly, groups with low $Q$ will be assigned group
masses that are systematically biased low. Such systematic biases
are difficult to control without additional, independent constraints
on the halo masses in which the group reside, and it is important
to be aware of these potential shortcomings.

\subsection{Purity and Completeness}
\label{subsec:purity_completeness}

Group finding is an archetypal example of a problem in which one attempts to identify a special sub-sample, a distinct galaxy group, from a larger population, all other galaxies.  For any such problem, two natural questions that arise are ``how contaminated is the sub-sample with incorrectly identified members?'', and ``how often do true members of the sub-sample fail to be included by the selection algorithm?''  The former question is conventionally quantified by the ``purity'' of the sub-sample, the latter by the ``completeness'' of the sub-sample.  In practice, there are various ways in which one could apply these two metrics to characterize the errors made by group finders.   

At first glance, the most straightforward approach to measure purity and completeness would seem to be to calculate each on a group-by-group basis.  This approach requires uniquely associating individual halos in the mock to individual groups in a group catalogue so that one may ask a question like ``what fraction of galaxies in a group {\em truly} belong to that group?'' (a measure of purity) or ``what fraction of galaxies in a halo are {\em correctly} assigned to a group?" (a measure of completeness).  Algorithmically, to measure completeness requires looking up a halo in the mock and tagging the galaxy members, then looking up the group associated with that halo in a group catalogue and counting how many of those galaxies were tagged as being members of the antecedent halo, relative to the number that were not tagged.  The measurement of purity is done by proceeding in the opposite direction.  This endeavour develops into a non-trivial task in the presence of fusing or fracturing, where multiple groups and haloes can be associated. 

In light of this, and our focus on the occupation statistics and not the properties of individual groups, we abandon a group-by-group approach in favour of a different one.  Instead, we consider the purity and completeness of our group catalogues focusing on the sub-samples of centrals and satellites, identification of which is one of the primary goals of a group finder.  In this case, it is no longer necessary to explicitly connect groups to haloes, as every galaxy has a unique mapping between central/satellite condition in the mock and central/satellite designation in a group catalogue.  Now, the appropriate question, e.g. for purity, becomes ``which fraction of satellites (or centrals) in a group catalogue are {\em truly} a satellite (or central) in the mock?''.  While this is a fundamentally different measure than the purity and completeness of the group memberships, we will show that the two questions are intimately related.   

We define the completeness of satellite galaxies identified by a group finder as:
\begin{equation}
C_{\rm sat}\equiv \frac{N_{\rm sat|sat}}{N_{\rm sat|sat}+N_{\rm cen|sat}},
\end{equation}
and the completeness of central galaxies as:
\begin{equation}
C_{\rm cen}\equiv \frac{N_{\rm cen|cen}}{N_{\rm cen|cen}+N_{\rm sat|cen}},
\end{equation}
where, for example, $N_{\rm sat|cen}$ is the number of galaxies identified as satellites but which are centrals in the mock.  In a complementary fashion, we define the satellite purity as:
\begin{equation}
P_{\rm sat}\equiv \frac{N_{\rm sat|sat}}{N_{\rm sat|sat}+N_{\rm sat|cen}},
\end{equation}
and the central purity as:
\begin{equation}
P_{\rm cen}\equiv \frac{N_{\rm cen|cen}}{N_{\rm cen|cen}+N_{\rm cen|sat}}.
\end{equation} 
Additionally, we briefly note that the total number of galaxies and inferred groups with this formalism is given by:
\begin{align}
N_{\rm gal} &= N_{\rm cen|cen}+N_{\rm sat|cen} + N_{\rm sat|sat}+N_{\rm cen|sat}, \\
N_{\rm groups} &= N_{\rm cen|cen}+N_{\rm cen|sat}.
\end{align} 
We plot the completeness and purity as a function of group mass for both centrals and satellites for each group finder in Fig. \ref{fig:purity_completeness}, where purity is a function of {\em assigned} group mass (purity calculations start by looking up galaxies in identified groups) and completeness is a function of {\em true} halo mass (completeness measurements begin by looking up galaxies in haloes).   

We can understand the physical significance of the various measures of purity and completeness by considering the effect of fusing and fracturing on each measurement.  Because centrals outnumber satellites, any fusing is likely to transform a central galaxy into a satellite.  This can be seen by considering the case of pure fusing during the group finding process (middle row in Fig. \ref{fig:artwork}).  This process will result in two {\em true} centrals being identified in one group\footnote{see appendix \ref{appendix:pernicious_effects} for an exceptional case}.  By definition, one of those centrals will be classified as a satellite in the resulting group.  Similarly, any fracturing is likely to transform a satellite into a central galaxy.  It follows that fusing generally decreases the purity (completeness) of satellites (centrals) and fracturing generally decreases the purity (completeness) of centrals (satellites).  While it is possible to cook-up circumstances where a combination of fusing and fracturing can result in no decrease in the purity and completeness measures (two groups exchanging centrals or satellites only), these appear to be sub-dominant failure pathways.

No group finder results in perfect purity or completeness of the central or satellite sub-samples.  This is the first indication that indeed some fusing and fracturing unavoidably takes place in the group finding process.  The {\tt YANG-GF} and {\tt TINKER-GF} groups show that $P_{\rm sat}$ and $C_{\rm sat}$ are generally lower than that of the centrals, becoming approximately equivalent at high mass.  This indicates that satellites are generally more difficult to identify.  This is the expected result for the identification of a sub-dominant population, a less extreme example of the perennial problem of locating a needle in a haystack.  On the other hand, we note that the {\tt BERLIND-GF} results in a particularly low satellite purity, with the benefit of producing a very high satellite completeness, $\sim 90\%$.  This is an indication that a significant amount of fusing is taking place with minimal fracturing in the group finding process.  The relative equivalence of the $P$ and $C$ measurements in the other two indicate that both processes are acting in near equilibrium.     

We note that all group finders show a trend of decreasing $P_{\rm cen}$ and $C_{\rm cen}$ towards increasing group mass.  At least one factor in this tend is due to the increasing group central inversion fraction for high mass haloes (shown in Fig. \ref{fig:inversion_fraction}).  This causes group finders to misidentify central galaxies, even in correctly identified groups.  In the limit where group membership is perfectly determined $P_{\rm cen}=1-f_{\rm cen,inv}$.  Thus, $f_{\rm cen,inv}$ sets an upper limit to $P_{\rm cen}$ that no group finder is likely to exceed.  That is, the algorithm to identify central galaxies in groups is fundamentally limited, and our measurements of $P_{\rm cen}$ and $C_{\rm cen}$ are simply reflecting this fact.   

Lastly, we define a closely related quantity to the purity and completeness measures, the satellite transition factor, $T_{\rm sat}$, a measure of the excess probability that a central galaxy in the mock is identified as a satellite galaxy by the group finder over the inverse process, as:
\begin{equation}
T_{\rm sat}\equiv \frac{C_{\rm sat}}{P_{\rm sat}} = \frac{N_{\rm sat|sat}+N_{\rm sat|cen}}{N_{\rm sat|sat}+N_{\rm cen|sat}}.
\end{equation}
If $T_{\rm sat} = 1$, then for each satellite that is misidentified as a central, a central is misidentified as a satellite ($N_{\rm sat|cen}=N_{\rm cen|sat}$).  If $T_{\rm sat}\neq 1$ then the inferred satellite fraction,
\begin{equation}
f_{\rm sat}^{\rm inf} = \frac{N_{\rm sat|sat}+N_{\rm sat|cen}}{N_{\rm gal}},
\end{equation}
will deviate from the true mock satellite fraction,
\begin{equation}
f_{\rm sat}^{\rm mock} = \frac{N_{\rm sat|sat}+N_{\rm cen|sat}}{N_{\rm gal}},
\end{equation}
where $f_{\rm sat}^{\rm inf}=T_{\rm sat}f_{\rm sat}^{\rm mock}$.  If more centrals transition to satellites, the inferred satellite fraction will increase, and vice versa.

\begin{figure}
  \includegraphics[width=\columnwidth]{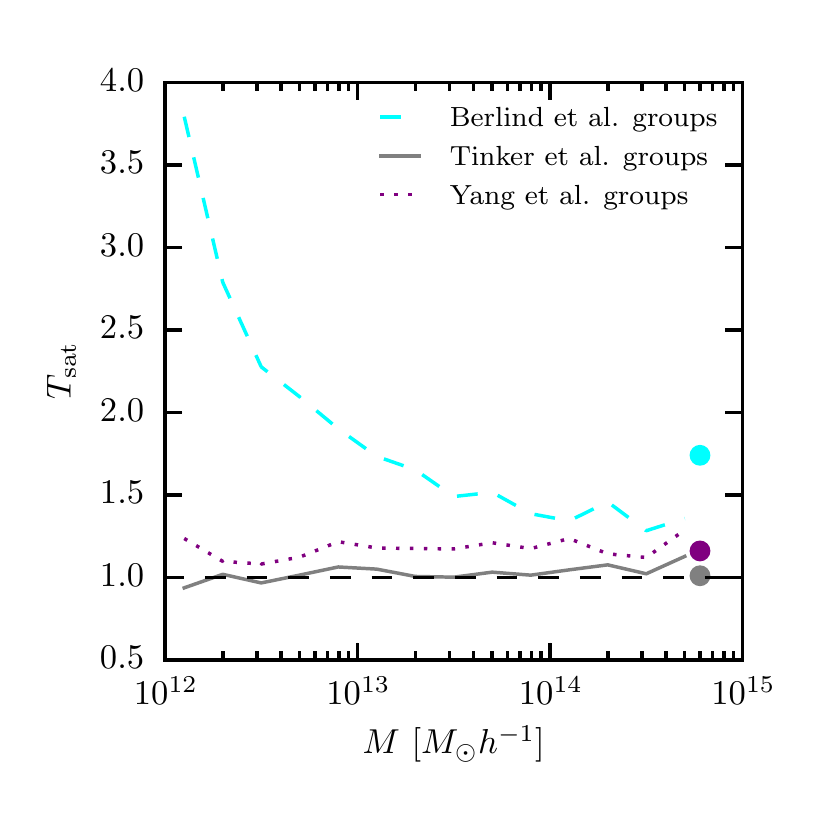}
  \caption{The satellite transition factor, $T_{\rm sat}$, as a function of group/halo mass for each group finder.  The filled circles are the value for the full sample.}
\label{fig:Tsat}
\end{figure}

We plot the value of $T_{\rm sat}$ as a function of group mass for each group finder in Fig. \ref{fig:Tsat}.  As expected from the purity and completeness measurements, the {\tt BERLIND-GF} has an elevated $T_{\rm sat}$, increasing for lower mass groups.  The {\tt YANG-GF} also displays a slightly elevated $T_{\rm sat}$, but it remains constant over the mass range considered.  From each of these observations, we expect the inferred satellite fractions to be increased for inferences from those two group catalogues (see \S \ref{subsec:results_fsat}).          

The difference in the value of $T_{\rm sat}$ for each group finder has its origins in the halo definitions and mocks used to tune the parameters of each group finder.  Group finders tuned on mocks using halo definitions that result in larger, more extended haloes relative to those in our mock will result in a group finder with a higher $T_{\rm sat}$, and vice versa.  We will return to this point in a discussion on the role of halo definitions in group finding in \S~\ref{subsec:discussion_halodef}.

\subsection{The Halo Transition Probability}
\label{subsec:HTP}

In this section we introduce a new statistic to quantify the errors made by group finding algorithms that encapsulates the effect of all three error modes discussed in this paper, in particular, capturing the correlated nature of the three.  The halo transition probability (HTP hereafter) measures the probability that a galaxy, with specified halo properties, transitions to a group, with specified group properties.  Given that the primary environmental dependence of our focus is halo mass, we formulate our new statistic in terms of the underlying halo mass and assigned group mass of individual galaxies.

Our formulation of the HTP measures the probability that a galaxy assigned to a group of mass $M_{\rm group}$ is truly located in a halo of mass $M_{\rm halo}$, i.e. $P(M_{\rm halo}|M_{\rm group})$.\footnote{This relation can also be inverted:  $P(M_{\rm group}|M_{\rm halo})= P(M_{\rm halo}|M_{\rm group})P(M_{\rm group})/P(M_{\rm halo})$}  For example, the HTP can answer the question ``what is the probability that a galaxy estimated to occupy a halo of mass $10^{14}M_{\odot}$ in a group catalogue truly occupies a $10^{12}M_{\odot}$ halo?''  However, this only gives an estimation of mass errors, one of the three primary goals of group finding.  In addition to halo mass, we are interested in the sub-populations of central and satellite galaxies.  So, we measure the transition probabilities of central/satellite designations in the group finding process, e.g. the probability that a central in the mock in a halo of mass $M_{\rm h}$ is identified as a satellite in the group catalogue given an identified group mass, $P({\rm sat_h}M_{\rm h}|M_{\rm g},\rm cen_g)$.  So, the functions that compose the HTP are:
\begin{align}
& P({\rm cen_h},M_{\rm h}|M_{\rm g},\rm cen_g) \\
& P({\rm sat_h},M_{\rm h}|M_{\rm g},\rm cen_g) \\
& P({\rm sat_h},M_{\rm h}|M_{\rm g},\rm sat_g) \\
& P({\rm cen_h},M_{\rm h}|M_{\rm g},\rm sat_g),
\end{align}
where the normalization is such that:
\begin{align}
\int \left[P({\rm cen_h},M_{\rm h}|M_{\rm g},\rm cen_g) + P({\rm sat_h},M_{\rm h}|M_{\rm g},\rm cen_g)\right]\mathrm{d}M_{\rm h} &= 1 \nonumber \\
\int \left[P({\rm sat_h},M_{\rm h}|M_{\rm g},\rm sat_g) + P({\rm cen_h},M_{\rm h}|M_{\rm g},\rm sat_g)\right]\mathrm{d}M_{\rm h} &= 1
\end{align}

The HTP can be thought of as a fingerprint of a specific group finder, characterizing the mapping between true properties of galaxies and the inferred properties in a group catalogue.  Each group finder will have a different HTP in detail, in as much as algorithms differ, and in general, the HTP will be dependent on the underlying mock.  That is, the response of the group finder will be dependent on the underlying model.  We will return to this point in a discussion of the prospects for using group catalogues in future studies of galaxy group properties in \S~\ref{subsec:discussion_correcting_gf}   

\begin{figure*}
  \includegraphics[width=190mm]{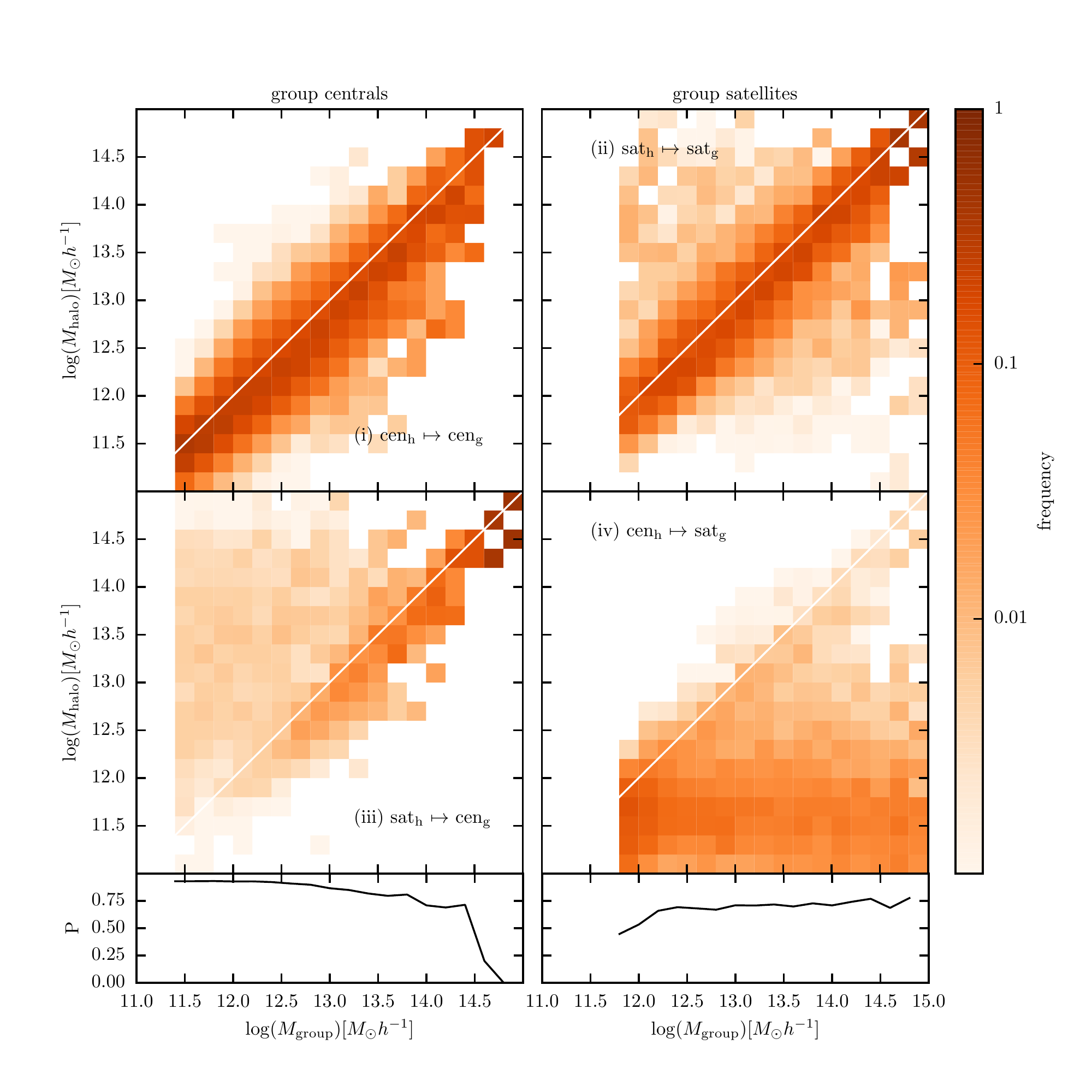} 
    \caption{Halo transition probability for the {\tt YANG-GF}.  The upper left quadrant(i) corresponds to central galaxies in the mock which were identified as central galaxies by the group finder, $P(M_{\rm h},\rm cen_h|{\rm cen_g, M_{\rm g}})$.  The upper right quadrant(ii) corresponds to satellite galaxies in the mock which were identified as satellite galaxies by the group finder, $P(M_{\rm h},\rm sat_h|{\rm sat_g, M_{\rm g}} )$.  The lower left quadrant(iii) corresponds to satellite galaxies in the mock which were identified as central galaxies by the group finder, $P(|M_{\rm halo},\rm sat_h|{\rm cen_g, M_{\rm g}})$.  The lower right quadrant(iv) corresponds to central galaxies in the mock which were identified as satellite galaxies by the group finder, $P(M_{\rm h},\rm cen_h|{\rm sat_g, M_{\rm g}})$.  The x-axis is the estimate of halo mass assigned using the output of the group finders and the y-axis is the halo mass from the mock.  The orange colour coding shows the frequency of a galaxies placed in 0.2 by 0.2 dex mass bins where the frequencies sum to one along the vertical axis at each group mass bin.  The lower panels show the purity of the centrals (left) and satellites (right) as a function of group mass.}
\label{fig:yang_HTP}
\end{figure*}

To show the morphology of the HTP, we plot an example, the {\tt YANG-GF} run over our mock, in Fig. \ref{fig:yang_HTP} (The HTP for the {\tt TINKER-GF} and {\tt BERLIND-GF} are qualitatively similar).  Considering the first quadrant, we describe the meaning of the HTP as follows: given that a galaxy is identified as a central in a group of mass $M_{\rm group}$, the probability that it is truly a central and resides in a halo of mass $M_{\rm halo}$ is given by the value of the HTP at $(M_{\rm g},M_{\rm h})$, and similarly for the remaining quadrants.    

The upper panels of the HTP in Fig. \ref{fig:yang_HTP}(i, ii) show that galaxies with the correctly identified central/satellite designation display a distribution of mock halo masses centred along the 1:1 line, i.e. the group finder can recover an unbiased estimation of the halo mass for these galaxies.  The bottom panels in Fig. \ref{fig:yang_HTP}(iii, iv) show incorrectly identified centrals and satellites.  Galaxies incorrectly identified as centrals, i.e. satellites in the mock, cluster in the upper-left region of the plot.  These galaxies are systematically assigned group masses that are below their true host halo mass.  This is a telltale signature of fracturing. Galaxies which are incorrectly identified as satellites, i.e. centrals in the mock, cluster in the lower-right region of the plot.  These galaxies are systematically assigned group masses that are above their true host halo mass.  This is a telltale signature of fusing.  Thus the morphology of the HTP is consistent with our picture of fusing and fracturing.

To show the relation of the HTP to the purity and completeness measures discussed in the previous section, in the bottom panels of Fig. \ref{fig:yang_HTP}, we plot the purity of centrals and satellites ($P_{\rm cen}$, $P_{\rm sat}$) as a function of $M_{\rm group}$, which involves an integration of the HTP. 

Another group finding error phenomenon the HTP illustrates clearly is the effect of misidentifying centrals because a satellite is the brightest group member.  In such a circumstance, one galaxy will likely transition from a satellite in the mock to a central in the group catalogue without affecting the estimate of group mass.  This is exactly what is shown by the HTP in the 1:1 line in the lower left quadrant(iii) and to a lesser extent in quadrant(iv).

By examining the HTP, two classes of colour-dependent errors should be expected.  The fusing mode in group identification is dominated by low mass centrals being assigned to higher mass groups and misidentified as a satellite.  Low mass central galaxies are likely to be blue compared to satellites in higher mass groups.  Thus, the resulting decrease in satellite purity should be expected to lower the inferred red fraction of satellites.  Conversely, the fracturing error mode is dominated by satellites in higher mass haloes being assigned to low mass groups and misidentified as centrals in the process.  The decrease in purity of the central galaxies is then likely to increase the red fraction of centrals.  We examine this specific result in \S~\ref{subsec:results_fred}. 

\section{Results}
\label{sec:results}

In this section we present our principal results pertaining to the fidelity with which group finders recover the true, underlying, statistical trends in our mock catalogues. We begin in \S~\ref{subsec:results_group_vs_halo} with a formal description of our technique for comparing the true statistical trends of the underlying galaxy distribution to those trends inferred by the group-finder(s).  In \S~\ref{subsec:results_hod} -  \S~\ref{subsec:results_conformity} we examine a set of occupation statistics: the HOD, satellite fraction, CLF, red fraction, and 1-halo conformity.

\subsection{Technique for Comparing Groups to Haloes}
\label{subsec:results_group_vs_halo} 

We have identified three general categories of group finding failures (see \S~\ref{sec:group_finding_errors}):
\ben
\item halo mass estimation errors
\item central/satellite designation errors
\item halo membership assignment errors
\een 
In the following analysis, we use these three types of errors individually, or in conjunction, to describe, in detail, how group finding algorithms introduce systematic errors in the measurements of various colour-dependent occupation statistics.

It is often difficult to disentangle these three error modes to identify the primary culprit responsible for the error in the inferred statistic.  It is particularly difficult to isolate the effect of membership errors, because membership errors generally result in errors in mass estimation and central/satellite designation.  However, in the absence of membership errors, it is easy to isolate the other two errors.  To assist in this task, we have created three versions of a ``perfect'' group finder, one where the group membership is taken directly from the mock (i.e. all galaxies which reside in a common halo are exclusively assigned to a common group), but where we have applied either (1.) our method of halo mass estimation, (2.) the central/satellite designation criteria, or (3.) both simultaneously.  Otherwise, all galaxy(-group) properties are taken directly from the mock.  In the first case, we can isolate the effect of our method to estimate group masses.  In the second case, we can isolate the effect of our method to identify the central satellite (or conversely the effect of a non-zero central inversion fraction in the mock), and in the third case we can see the combination of these two effects in the absence of membership errors.  We show the results of the ``perfect'' group finder(s) to further clarify, when appropriate, which error modes are important for a statistic.  In some cases it is not possible to decouple these error modes, and an appeal to all three working in concert must be made to explain the observed error.  

In particular, throughout \S~\ref{sec:results} we formulate our comparisons as follows. First, we choose some particular statistic of the galaxy distribution, e.g., $\langle L_{\mathrm{cen}} \rangle$, and use our mock galaxy catalogue to measure the true dependence of this statistic on halo mass, $M_{\rm halo}$.  Second, we repeat this exercise, but instead use the group catalogue to infer the dependence of the same statistic on $M_{\rm group}$.  Comparing the inferred and true dependence allows one to assess how effectively group catalogues can be used to directly measure the statistical relationship between galaxy(-group) properties and halo properties.  All error bars are calculated by 50 bootstrapped group samples from the resulting group catalogues.

\subsection{The HOD}
\label{subsec:results_hod}

We begin our discussion of recovered statistics with the HOD.  In the HOD formalism, the occupation statistics of galaxies are encoded by $P(N_{\mathrm{gal}}|M_{\rm halo}),$ the probability that a halo of mass $M_{\rm halo}$ hosts $N_{\mathrm{gal}}$ galaxies satisfying some sample selection criteria, such as a brightness and/or a colour cut.  In this section, we study how effectively one may use group catalogues to directly infer the first moments of the HOD, namely  $\langle N_{\mathrm{gal}}(M_{\rm hlao}) \rangle$.

\begin{figure*}
    \includegraphics[]{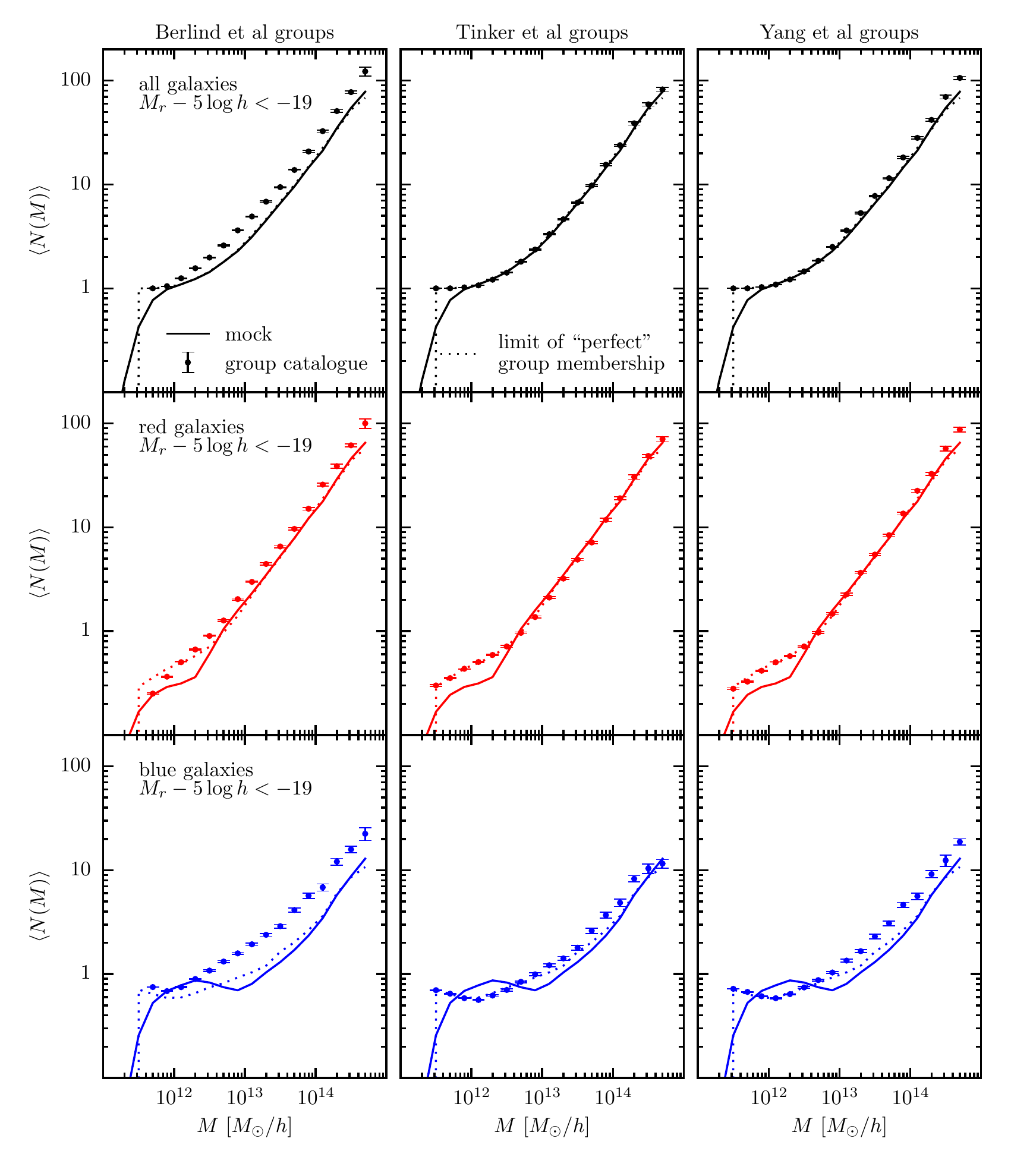} 
    \caption{The HOD for the group finders run over the mock.  The top row is for all galaxies in the mock, the middle for red galaxies, and the bottom for blue galaxies. The left column is for the{\tt BERLIND-GF}, the middle for the {\tt TINKER-GF}, and the right for the {\tt YANG-GF}.  Dark lines are taken from the mock, and points with error bars are the group finder results.  The dotted line is the result of a ``perfect'' group finder, where halo masses are estimated by abundance matching on total group luminosity.  The error bars are calculated from 50 bootstrapped samples of the group catalogues.}
\label{fig:HOD}
\end{figure*}

The results of the direct measurement of the HOD from the group finders (points) and the relation taken directly from the mock (solid lines) are shown in Fig. \ref{fig:HOD}.  The HOD was measured for three samples, the volume limited $M_r-5\log h<-19$  (top row), red sub-sample (middle), and blue sub-sample (bottom).  This was done for each group finding algorithm: {\tt BERLIND-GF} groups (left column), {\tt TINKER-GF} groups (middle), and {\tt YANG-GF} groups (right). 

 \begin{figure*}
   \includegraphics[]{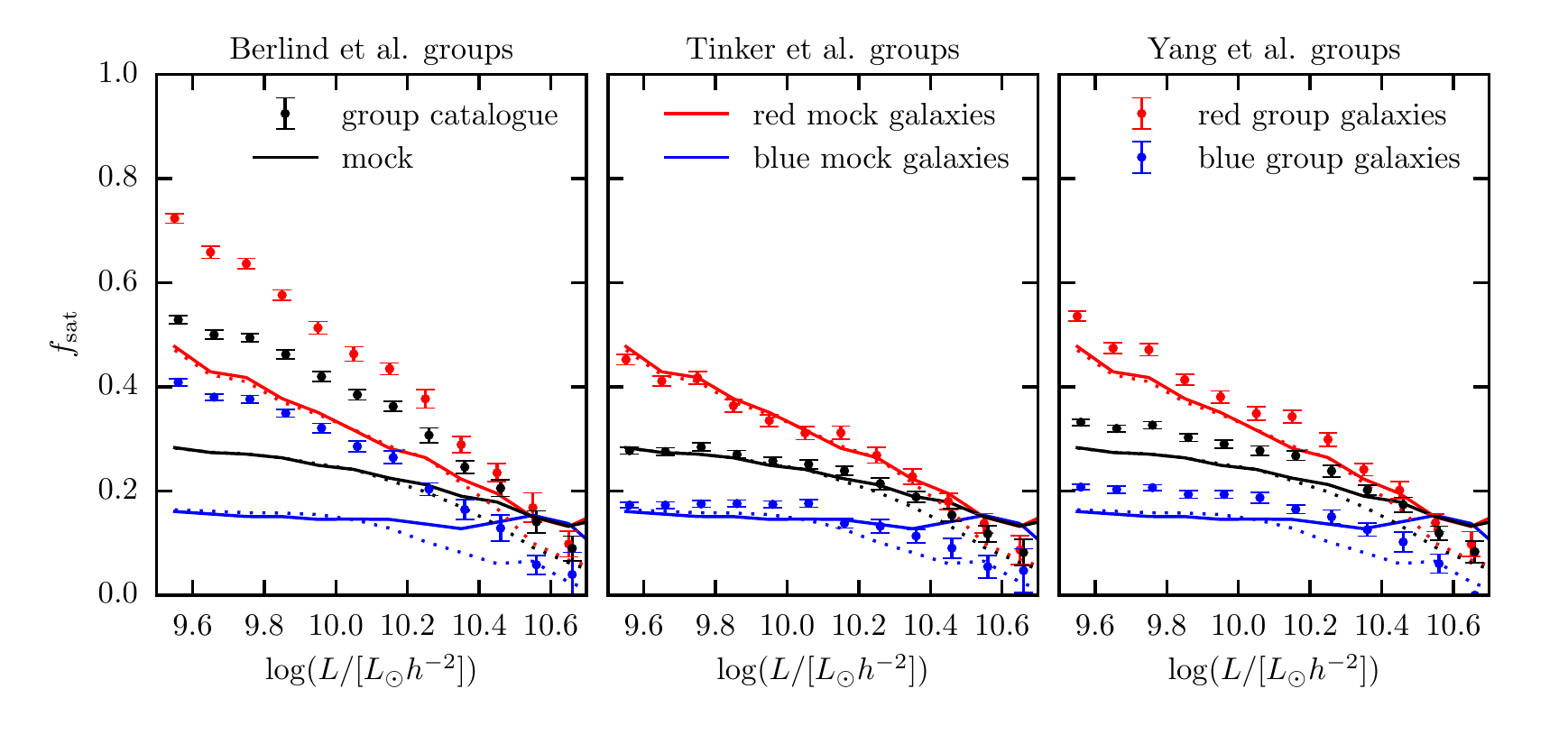} 
   \caption{The satellite fraction as a function of galaxy luminosity for red, blue, and all galaxies compared to the mock for each group finder: Berlind et al groups (left), Tinker et al groups (middle), and Yang et al groups (right).  Results for the full sample of galaxies is shown in black, the red sub-sample in red, and the blue sub-sample in blue.  The group finder results are shown as points with error bars, while the intrinsic mock halo level statistics are shown as solid lines.  The result of a ``perfect'' group finder (where group membership is equivalent to that in the mock) is shown as dotted lines.  The inferred statistic uses only the satellite/central determination from the group finders.}
\label{fig:sat_fraction}
\end{figure*}  

An organizing principle for much of what follows: is that {\em group finder errors tend to equalize the properties of distinct galaxy sub-populations}.  Using this intuition, we can explain the specific effects seen in the HODs measured directly using the three group finders.  Focusing initially on the HODs for the full galaxy sample (shown in the top row of Fig. \ref{fig:HOD}), we note two important effects. The inferred transition from $\langle N_{\rm gal}(M)\rangle = 0$ to $\langle N_{\rm gal}(M)\rangle = 1$ is too sharp.  This is not a failure of the grouping algorithms \emph{per se}, but an unavoidable limitation of our implementation of abundance matching to estimate $\mgroup$.  This can be seen by examining the dashed line, representing the result using a ``perfect'' group finder that uses our implementation of halo mass estimation.  As discussed in the section on halo mass estimation errors, abundance matching with no scatter on total group luminosity will result in a HOD with a step function at the low mass end, a feature that is not present in our mock.  

Also apparent in the full sample HOD for the {\tt BERLIND-GF} groups is an overestimation of $\langle N_{\mathrm{gal}}|M \rangle$ at intermediate to high masses.  This is a consequence of the {\tt BERLIND-GF} over-fusing groups relative to our mock, resulting in a high satellite transition factor, $T_{\rm sat}\simeq1.75$.  This results in the satellite fraction being too high, and an overestimate of the $\langle N_{\rm gal}|M \rangle$.  The {\tt TINKER-GF} and {\tt YANG-GF} have  $T_{\rm sat}\simeq1$ and do not exhibit this same error, or at least to the same magnitude.  We stress that $T_{\rm sat}>1$ for the {\tt BERLIND-GF} groups is likely a consequence of how the group finder was tuned, particularly the mismatch between the FoF halo galaxy populations it was tuned to reproduce, and the SO halo galaxy populations present in our mock.  We leave a more thorough discussion of the role of halo definitions in group finding to \S~\ref{subsec:discussion_halodef}.

The colour split HODs, red galaxies (middle row) and blue galaxies (bottom) in Fig. \ref{fig:HOD}, show further effects that appear to be a more general result of all three grouping algorithms.  Principally, each group finder displays an inability to capture the shape of the colour-dependent HOD for the region dominated by central galaxies at the low halo mass end, and all group finders predict far too many blue galaxies in the region dominated by satellite galaxies in higher mass haloes.  These two errors originate in two distinct limitations of group finders, namely halo mass estimation errors in the region dominated by central galaxies, and membership errors in the region dominated by satellites. 

Abundance matching on total group luminosity to estimate halo masses for groups with $\langle N_{\rm gal}\rangle \simeq 1$ is not able to capture the complex relation between galaxy colour and halo property seen in our mock.  That is, in the mock there is a correlation between halo mass and colour at fixed luminosity, so an estimate based on only luminosity misses this correlation.  The result of a perfect group finder that uses the halo mass estimation algorithm is shown as dashed line, perfectly matching the {\tt YANG-GF} and {\tt TINKER-GF} results at low halo masses.  Again, this indicates that using group luminosity to assign halo mass estimates for groups with $\langle N_{\rm gal}\rangle \simeq 1$ cannot altogether capture complex relations at fixed luminosity (see \S ~\ref{subsec:mass_errors} for a more detailed theoretical discussion of this issue). 

The error in the regime where $N_{\rm gal}>1$ has an all together different origin.  Membership errors will generally result in impurities in the satellite population.  Because most galaxies are central galaxies, which are on average bluer than satellite galaxies, at a fixed luminosity, fusing of central galaxies into larger groups will result in an increased number of blue centrals identified as satellite galaxies in groups.  Similarly, fracturing of groups is likely to result in incompleteness in the red satellite population, resulting in a decreased number of red satellites in groups.

These trends appear in the inferred HOD measured by the group finders.  All three group finders infer an increased number of blue galaxies in high mass haloes as a result of fusing induced impurity in the satellite sample.  However, the inferred number of red galaxies is much closer to the true value for all three group finders, with subtle differences.  The group finder with the highest satellite completeness, {\tt BERLIND-GF}, infers more red galaxies in these haloes than the group finder with the lowest satellite completeness, {\tt TINKER-GF}.

\subsection{Satellite Fractions}
\label{subsec:results_fsat}

The satellite fraction of galaxies, meeting a particular selection criterion, is an important quantity for many studies including galaxy-galaxy lensing, the pairwise velocity dispersion of galaxies, cluster finding, and the large scale clustering of galaxies.  Group catalogues provide a way to directly infer this quantity.  Here we measure the red and blue satellite fraction as a function of galaxy luminosity, $f_{\rm sat}(L|\rm red)$ and $f_{\rm sat}(L|\rm blue)$.  Note that this statistic does not make use of the halo mass estimation of groups. 

In Fig. \ref{fig:sat_fraction} we plot the satellite fraction for red and blue galaxies from our mock (solid lines) and as measured by the group finders (points).  Also, we show, as a dashed line, the result of a ``perfect'' group finder where the satellite/central designation is determined using the same criteria as the group finders.

{\tt BERLIND-GF} groups infer a satellite fraction that is high relative to the mock as one would expect given the $T_{\rm sat}\simeq1.75$.  The excess fusing affects both the red and blue galaxy samples.  Here we can also see that the {\tt YANG-GF} groups over-estimate the satellite fraction to a lesser degree, affecting both red and blue galaxies.  The {\tt TINKER-GF} groups recover the statistic very accurately at all but large luminosities, where satellite inversion becomes important 

At first glance, it is surprising that the group finder inferred satellite fraction is not significantly under-estimated for red galaxies and over-estimated for blue galaxies.  However, because age-matching results in 2-halo conformity, the neighbouring galaxies to any group in the mock are more similar in colour to the group members than the average galaxy in the mock.  Because of this, any membership error will produce less of an error in a colour-dependent statistic than if a random galaxy had been merged into a group.  This is how each group finder can have a significant amount of fusing and fracturing while keeping the relative number of red to blue satellites approximately equal.  This is a specific prediction of age-matching, and a mock that does not have any 2-halo conformity will generally result in a larger colour-dependent error than the one measured here.  We will come back to this point in \S~\ref{subsec:discussion_quenching}.

\subsection{The CLF}
\label{subsec:results_clf}

\begin{figure*}
    \includegraphics[]{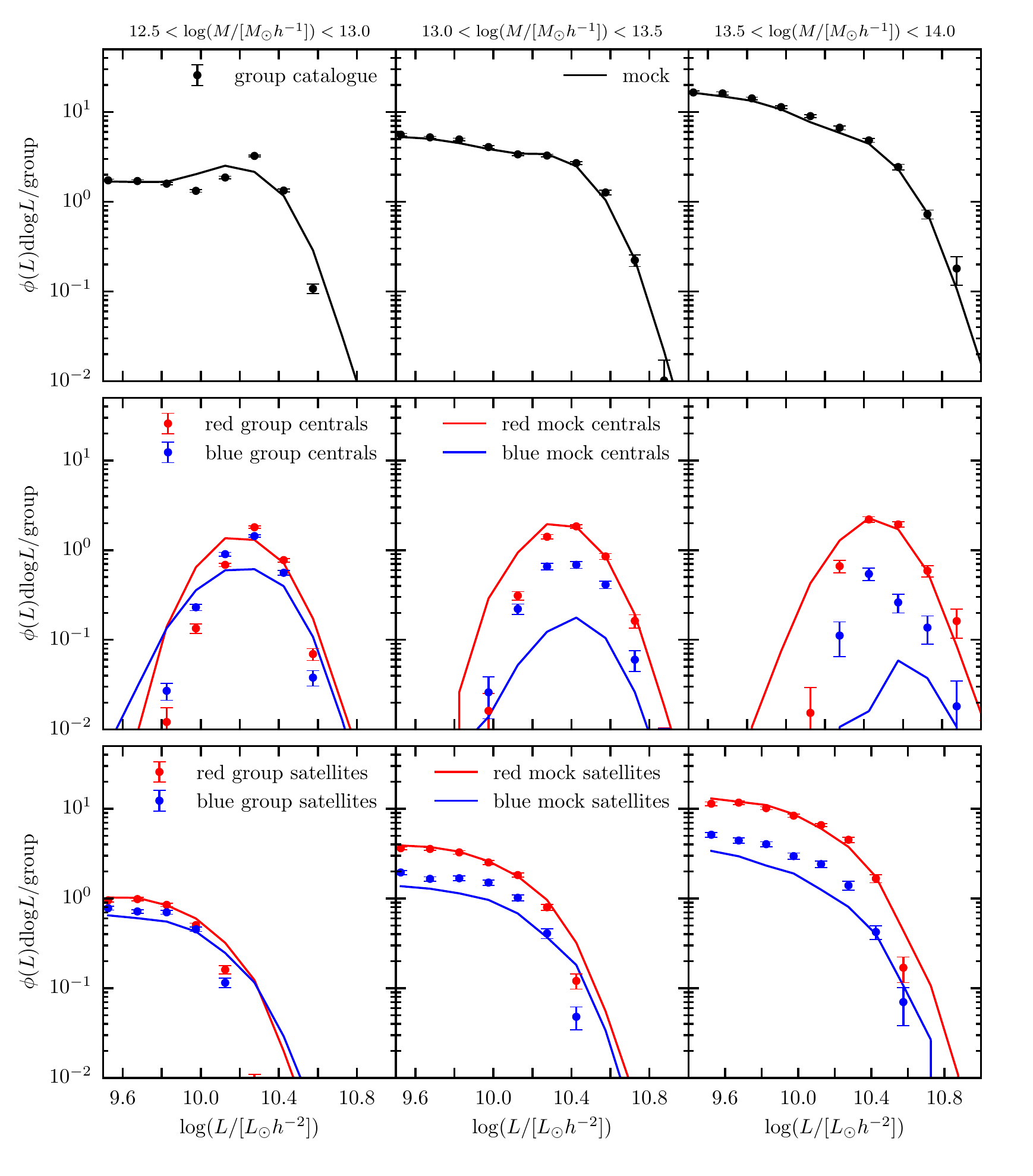} 
    \caption{Conditional luminosity functions (CLF) for the {\tt YANG-GF} run over our mock. The group finder results are shown as points with error bars. The intrinsic mock results are shown as lines.  First row: the black line and points are for the full halo/group population.  Second row: the dashed red line and red points with error bars are for red central galaxies only.  the blue lines and blue  points with error bars are for blue central galaxies only.  Bottom row: the red line and red points with error bars are for red satellites only.  The blue line and blue points with error bars are for blue satellites only.}
\label{fig:clf}
\end{figure*}   

\begin{figure*}
   \includegraphics[]{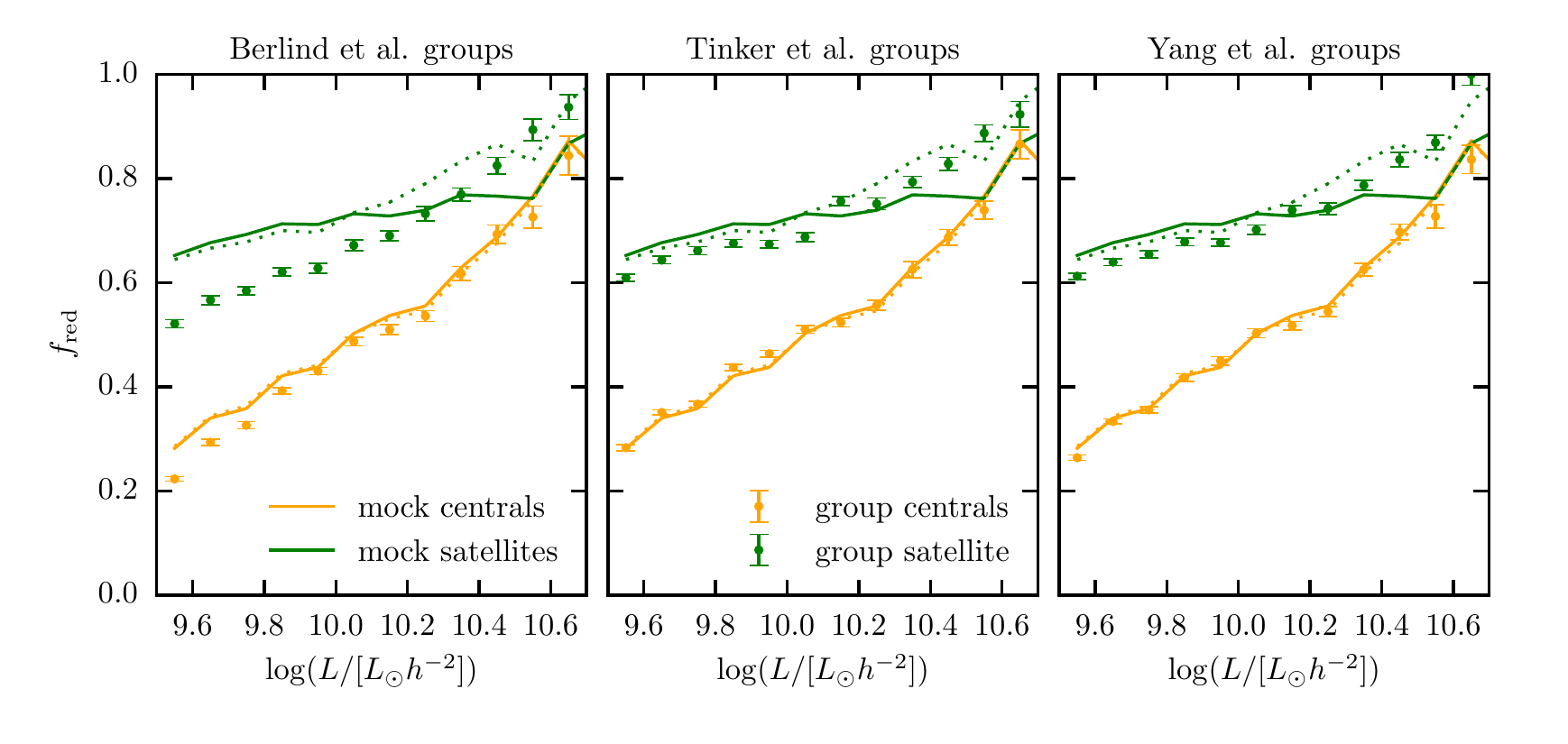} 
   \caption{The inferred red fraction of galaxies as a function of galaxy luminosity for centrals and satellites compared to the mock for each group finder: Berlind et al groups (left), Tinker et al groups (middle), and Yang et al groups (right).  Results for centrals are shown in orange and satellites in green.  The group finder results are shown as points with error bars, while the intrinsic mock halo level statistics are shown as solid lines. The results of a ``perfect'' group finder (where group membership is equivalent to that in the mock) is shown as dotted lines.  The inferred statistic uses only the satellite/central determination from the group finders.}
\label{fig:f_red_L}
\end{figure*} 

The conditional luminosity function (CLF herafter), $\phi(L|M_{\rm halo})\mathrm{d}L$, is the average number of galaxies in a luminosity range, $L+\mathrm{d}L$, that occupy a halo of mass $M_{\rm halo}$.  Group catalogues, in principle, provide a manner in which the CLF can be directly measured \citep[e.g.][]{Yang:2008tg}.  For our measurements of the CLF, we have made use of all properties of the group finding algorithms: membership determination, halo mass estimation, and central/satellite designation.  Therefore, the CLF measurements are the first statistic discussed so far which in principle could suffer from any combination of all three general group finding errors: membership misidentification, halo mass estimation errors, and central/satellite designation errors.   

In Fig. \ref{fig:clf} we show the CLF in three mass bins (columns) as measured by the {\tt YANG-GF}.  The full CLF (top row) has been split into a central (middle row) and satellite component (bottom row), and further split into red (red points/lines) and blue (blue points/lines) sub-samples of the latter two components.  Largely, the CLF is recovered remarkably well for the non-colour split and the red sub-sample at all three halo masses considered here.  In detail for the errors that are made, we see a similar effect in the CLF measurements as in previously discussed statistics, namely that group finder errors tend to equalize the properties of distinct galaxy sub-populations.  In this case, the group finder inferred measurements of the central galaxy component of the CLFs are more narrow, and the blue and red populations are inferred to be more equal than those in the mock.  This is a result of designating the brightest group member the central galaxy.  A group where the central has been misidentified is likely to be one where the central has a relatively low luminosity.  Thus this algorithm to identify centrals will generally prematurely truncate the distribution of low central luminosities and consequently result in a more peaked distribution (the number of centrals per group is conserved).  The corollary to this is that the satellite component falls off more rapidly towards higher luminosities in the group finder inferred statistic (which we also see).

The satellite component is most affected by membership determination errors resulting in impurities and incompleteness in the blue and red satellite sub-samples.  Groups that have undergone fusing receive boosts in the number of blue satellites, and fracturing is most likely to result in a red satellite being identified as a central.  This affects the measurement of the CLF most strongly by overestimating the luminosity function for the sub-dominant blue satellites, bringing the measurement closer to that of the red satellites.

\subsection{Red Fractions}
\label{subsec:results_fred}

\begin{figure*}
    \includegraphics[]{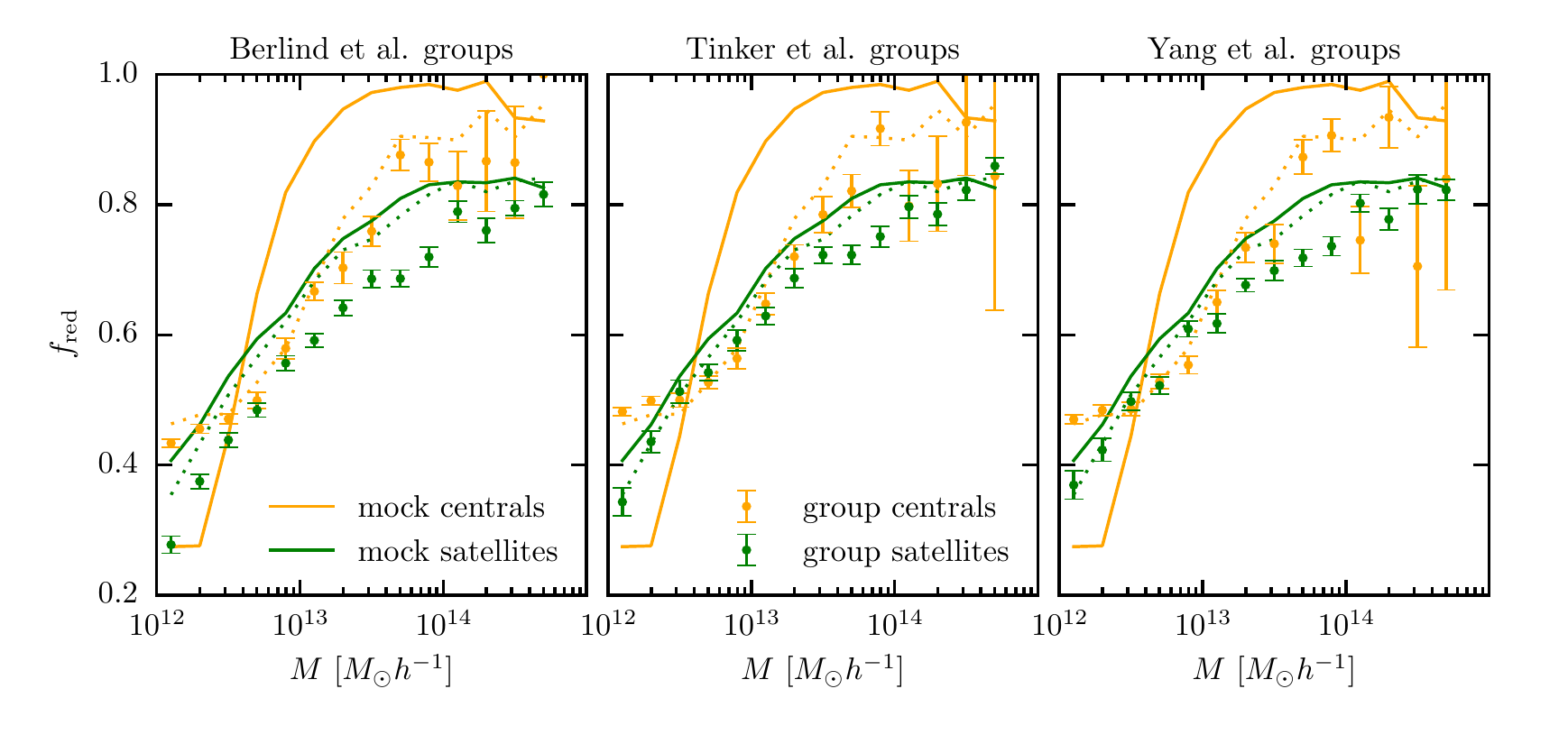} 
    \caption{The inferred red fraction of galaxies as a function of group/halo mass for central galaxies and satellite galaxies compared to the mock for each group finder: Berlind et al groups (left), Tinker et al groups (middle), and Yang et al groups (right).  Results for centrals are shown in orange and satellites in green.  The group finder results are shown as points with error bars, while the intrinsic mock halo level statistics are shown as solid lines. The results of a ``perfect'' group finder where halo masses and central/satellite designation are determined in the same way as the group finders and where group membership is equivalent to that in the mock is shown as dotted lines.  The inferred statistic uses the satellite/central determination and halo mass estimation from the group finders.}
\label{fig:f_red_M}
\end{figure*}

The red fraction of satellites and centrals provides an important constraint on galaxy evolution models \citep[e.g.][]{vandenBosch:2008fv, Peng:2012bp, Wetzel:2014up}.  In principle, group finders provide a method to directly measure this statistic.  In this section we discuss the recoverability of $f_{\rm red|cen}$ and $f_{\rm red|sat}$ as a function of galaxy luminosity (see Fig. \ref{fig:f_red_L}) and halo/group mass (see Fig. \ref{fig:f_red_M}) by group finding algorithms.

First, we examine the red fraction as a function of galaxy luminosity in Fig. \ref{fig:f_red_L}.  The inferred red fraction of satellites and centrals tend to converge towards a common value, where, in general, the red fraction of satellites, $f_{\rm red|sat}(L_{\rm gal})$, is under-estimated, and the red fraction of centrals, $f_{\rm red|cen}(L_{\rm gal})$, is over-estimated.  As expected from an examination of the CLF and satellite fractions, at high luminosities, the red fraction of satellites is over-estimated relative to the mock where central inversion becomes important.  Interestingly, the {\tt BERLIND-GF} measurements under-estimate the red fraction of centrals.  This is strongly affected by the 2-halo conformity present in the age-matching mock.  Because the {\tt BERLIND-GF} groups suffer a large amount of fusing, a significant population of centrals are merged into groups.  The centrals that are likely to merge into larger groups are those that live in more dense environments, and the remaining low luminosity centrals that were not merged are more likely to be blue.  This phenomena also affects the {\tt YANG-GF} and {\tt TINKER-GF} groups where the under-estimate of the satellite red fraction is significantly reduced and there is almost no error in the inferred red fraction of centrals.            

In contrast, the $f_{\rm red|sat}(M_{\rm group})$ and $f_{\rm red|cen}(M_{\rm group})$ statistics (Fig. \ref{fig:f_red_M}) display a more interesting behaviour in the deviation from the mock value.  The satellite red fraction continues to be underestimated by the group finders, consistent with previous measurement, but the measured central red fraction deviates significantly from the true underlying behaviour in the mock.  In particular, the sharp transition between low and high central red fraction at $\sim 10^{13} ~ M_{\odot}$ is completely washed out by the group finding algorithms.  This is primarily be a consequence of mass estimation errors at low and intermediate masses and central identification errors at high masses and not effects from membership determination errors--all three group catalogues produce nearly identical measurements of the central red fraction as a function of group mass.  This may have important implications for the inference of quenching mass scales, and we will return to a discussion of the consequences of this result in \S~\ref{subsec:discussion_quenching}

\subsection{1-Halo Conformity}
\label{subsec:results_conformity}

A primary motivation of this paper is to investigate the ability of group catalogues to study the galactic conformity phenomena. 1-halo galactic conformity was originally detected using the {\tt YANG-GF} in \cite{Weinmann:2006hu}.  Consistent with that detection, we define 1-halo conformity as the tendency of red central galaxies to host redder satellites than blue centrals at a {\em fixed halo mass}.  Galactic conformity serves as a strong test of galaxy evolution and formation models as it violates the HOD formalism.  Age-matching naturally gives rise to 1-halo conformity, and our mock is ideally suited to test the sensitivity of group finders to conformity.     

However, it is also desirable to test the group finders on a mock which contains no conformity in order to ascertain if measurements inferred from group finders have a tendency to induce a false conformity measurement.  For this purpose, we create a shuffled version of the age-matching mock which preserves the HOD, but wherein the conformity phenomenon is removed by shuffling centrals and satellites between haloes of equal mass, destroying the central-satellite colour correlation at fixed halo mass (see Appendix \ref{appendix:shuffling} for details on how this mock was created).

\begin{figure*}
	\includegraphics[]{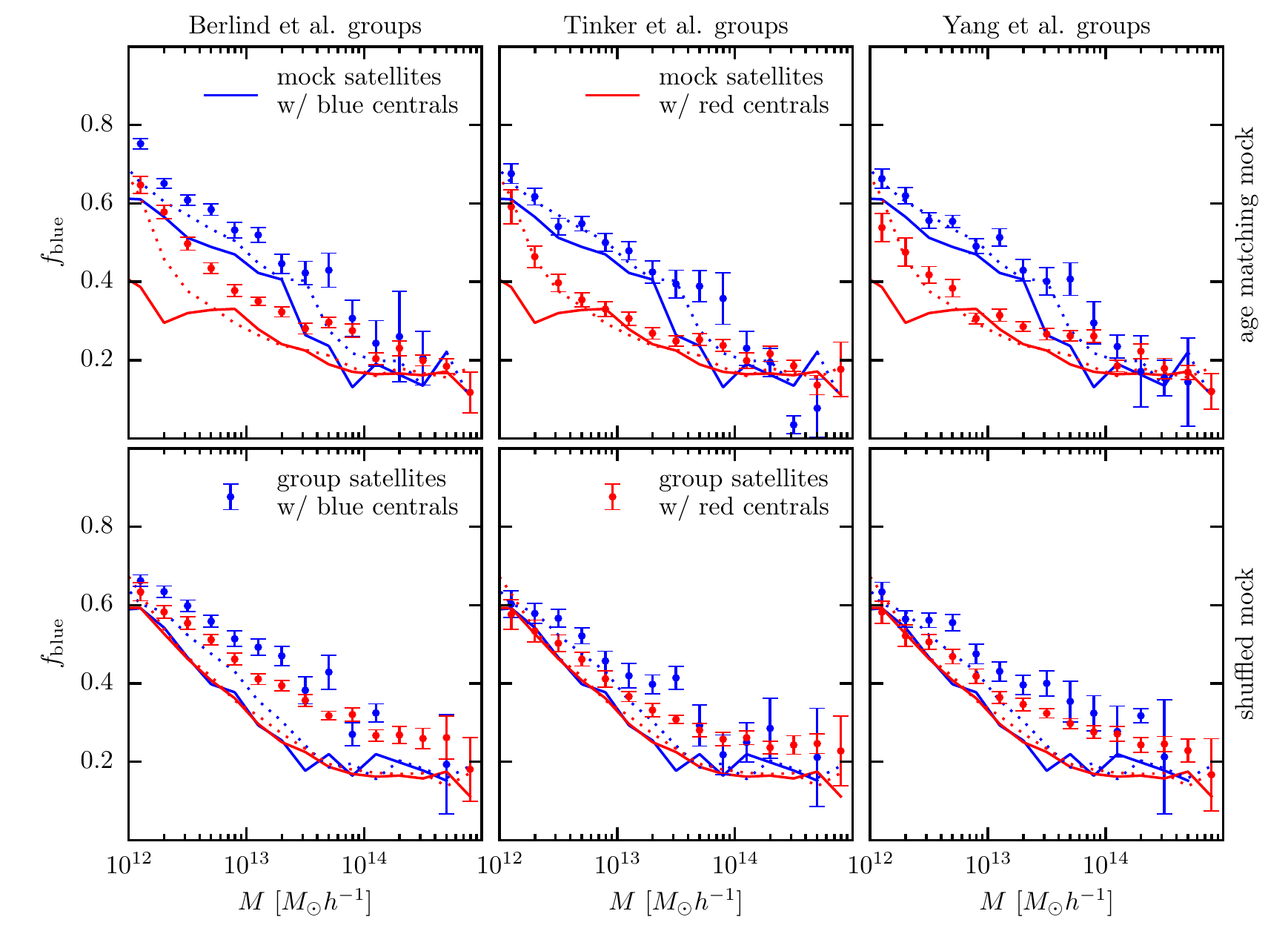} 
	\caption{1-halo galactic conformity measured in the age-matching mock (top) and the shuffled mock(bottom) for the {\tt BERLIND-GF} groups (left), {\tt TINKER-GF} groups (middle), and {\tt YANG-GF} groups (right).  The blue fraction of satellites is plotted against halo/group mass for groups/haloes with a red central galaxy (red) and blue central galaxy (blue).  The intrinsic mock halo statistics are shown as lines, and the group finder results are shown as points with error bars.  The dotted lines are the result of a ``perfect" group finder.  The conformity ``signal'' is the separation between blue lines (points) and red lines (points).  No separation implies no 1-halo conformity, and conversely, separation implies 1-halo conformity.}
\label{fig:conformity}
\end{figure*}

In Fig. \ref{fig:conformity} we plot the blue fraction of satellites in haloes (groups) with red central galaxies as red lines (red points) and those with blue central galaxies as blue lines (blue points).  This is done for each group finder in the three columns for both the age-matching mock, which contains conformity in the top row, and for the shuffled mock, which does not contain conformity, in the bottom row.  The conformity ``signal'' in these plots is the separation between blue lines (points) and red lines (points).  

Immediately, we conclude that all three group finders are sensitive to the existence of 1-halo conformity in the age-matching mock; however, all three group finders also show a smaller, but still significant amount of conformity in the shuffled mock, where there is no intrinsic conformity.  First, we consider the measurements on the mock with conformity.  The group finders are generally successful at recovering this particular measure of 1-halo conformity in terms of recovering the magnitude of the $f_{\rm blue}$ separation between between red and blue centrals; however, the over-all normalization of the measured blue fraction is slightly high.  At the lower mass end, $M<\sim10^{12.5}M_{\odot}$, the blue fraction of galaxies around red centrals is significantly over-estimated, eventually resulting in the inferred conformity signal going to nearly zero. We attribute this trend to a limitation in the halo mass assignment method.  To demonstrate this, in Fig. \ref{fig:conformity} we have plotted the result of a perfect group finder as dotted lines where we have used the total group luminosity to estimate group mass.  We will return to this mass error point when describing the effects seen in the shuffled mock.  

Now, we turn our attention to the measurement on the shuffled mock.  All three group finders show an elevated blue satellite fraction for all masses compared to the mock.  In addition, all three group finders show an increased blue satellite fraction around blue central galaxies at fixed inferred group mass for $M_{\rm group}<\sim 10^{14}M_{\odot}$ relative to that inferred around red centrals (induced conformity).  The first effect, the overall elevated blue satellite fraction, is a repercussion of the removal of 2-halo conformity in the shuffled mock.  It is the presence of 2-halo conformity in the age-matching mock which reduces the effect of fusing and fracturing on the inferred red and blue satellite fractions discussed in previous sections.  Without this mitigating effect, the colour errors are generally larger.  

There are two effects which contribute to the measured conformity signal in the shuffled mock.  First, because the red fraction of central and satellite galaxies increases with halo mass, a false 1-halo conformity measurement can be induced when groups are subject to either fracturing or fusing (any membership failure will result in one or both of these failure modes being present). Consider a group that has been purely fractured--its parts will systematically be assigned a halo mass below its members' true halo mass, resulting in both the assigned central and satellite populations being redder than correctly identified groups of the same assigned halo mass.  Conversely, purely merged groups' members will be assigned a halo mass that is larger than its members' true halo mass, resulting in a group with bluer centrals and satellites relative to correctly identified groups of the same halo mass (examine the galaxy colours in Fig. \ref{fig:artwork}).  From this, we conclude that any membership errors generally result in an induced or elevated conformity signal.  

\begin{figure}
	\includegraphics[]{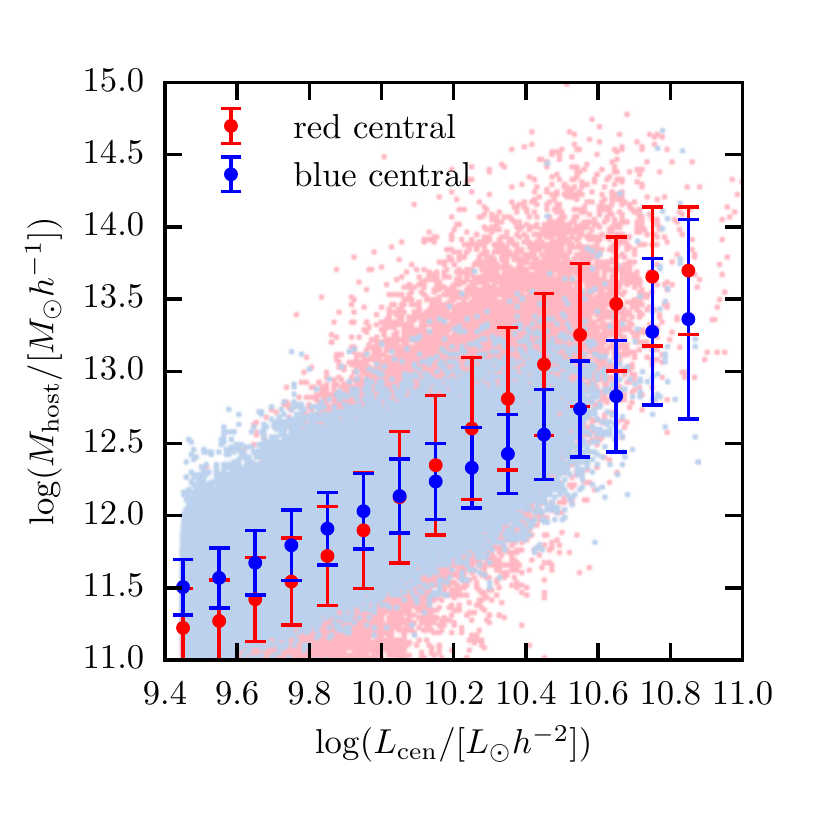} 
	\caption{mock halo mass vs. mock central luminosity for red centrals (red points and markers with error bars indicating the scatter) and for blue centrals (blue points and markers with error bars indicating the scatter).  At fixed central luminosity, blue centrals occupy lower mass haloes than red centrals for haloes more massive than $\sim 10^{12}M_{\odot}$ (and vice versa for lower mass haloes).  In the limit where group luminosity is determined by central luminosity, at equal assigned group mass, groups with a blue central will truly be associated, on average, with a lower mass halo than groups with a red central.}
\label{fig:Lcen_Mhalo}
\end{figure}

Second, the relative over (under-)estimation of the satellite blue fraction could have been expected around blue (red) centrals for our mocks given the relation between galaxy colour and halo mass at fixed luminosity.  In Fig. \ref{fig:Lcen_Mhalo} we show the halo mass-central luminosity relation for red and blue centrals in the age-matching mock (which is equivalent in the shuffled version).  This phenomenon occurs in our luminosity based mock because at fixed $V_{\rm peak}$ there is a non-random correlation between halo age and mass.  Estimating group mass using only luminosity results in a colour dependent mass error because the scatter in the $M_{\rm halo}-L_{\rm group}$ relation is correlated with galaxy colour (see the last paragraph of \S~\ref{subsec:mass_errors} for a more theoretical discussion).  Specifically, in haloes more massive than $10^{12.5}M_{\odot}$, where the satellite fraction begins to become significant for our sample, at a fixed central luminosity, blue centrals occupy lower mass haloes than red centrals.   At a given assigned $M_{\rm group}$, groups with blue centrals are likely to truly occupy lower mass haloes than groups with red centrals, and because more massive haloes have redder satellite populations than lower mass haloes, this systematic mass estimation error results in an induced conformity signal.  We show the magnitude of this error in the conformity signal measured using a perfect group finder as dotted lines in Fig. \ref{fig:conformity}.  This is the same effect, but in the opposite direction, which is responsible for the overestimation of the satellite blue fraction around red centrals in the age-matching mock at low masses (top panel).  Put another way, correlation between galaxy colour and halo mass at fixed luminosity can induce a conformity signal or alter the magnitude of the signal.       

\section{Discussion}
\label{sec:discussion}

\subsection{Implications for Group Catalogue Studies}

\subsubsection{Central and Satellite Quenching}
\label{subsec:discussion_quenching}

It is known from observations that quenched galaxies are more likely to have larger stellar masses and live in higher galaxy over-densities than star forming galaxies.  Developing a model which reproduces the quenched fraction of galaxies as a function of environment and luminosity/stellar mass remains a challenge for galaxy formation models.  Galaxy group catalogues provide a valuable resource to study this phenomena, allowing one to isolate the effect of environment and distinguish between central and satellite populations 

As we have seen, galaxy group catalogues tend to equalize the properties of distinct galaxy populations, with galactic conformity being a subtle exception.  This becomes an important bias to understand when comparing the occupation dependence on halo mass of star forming and quenched galaxies.  We have shown that group finders tend to decrease the inferred red fraction of satellites, while increasing it for central galaxies.  Furthermore, by examining this statistic using various shufflings of our mock (e.g. the `shuffled' mock described in Appendix \ref{appendix:shuffling}) we see that the magnitude of the error is dependent on the underlying model.

The quenched fraction of satellites and centrals (or number) as a function of halo mass has been measured \citep{Yang:2008tg,Yang:2009cm, Weinmann:2006hu, Weinmann:2009gf, Wetzel:2012lk, Peng:2012bp, Knobel:2014tx} using group catalogues without accounting for systematic errors associated with making these measurements.  Instead group finders should be forward modelled, by running the group finder over galaxy mocks, to make fair comparisons to observations \citep[e.g.][]{Skibba:2011dx,Wetzel:2013dw,Hearin:2013km,Watson:2015gq}.  We see that for this particular statistic, the dominate source of systematic error in our inference comes from correlated scatter in galaxy colour at fixed central luminosity and central inversion in our mock.  We leave a thorough study of the inference of central quenching statistics to a forthcoming paper.  

Even if the error for a relevant measure in our study is small, it is worth noting that this is significantly influenced by 2-halo conformity.  We believe 2-halo conformity is a realistic phenomenon to include, and that for the purposes of this study is accurately included in our mock, but, it may be a poor crutch to rely upon to correct for group finding errors, especially for galaxy properties other than colour or sSFR.  We can estimate the importance of this effect for the satellite red fraction shown in \S~\ref{subsec:results_fred} as follows--if group membership errors are random (galaxies that are mislabelled as satellites are drawn from the total population of centrals), in terms of purity and completeness (see \S~\ref{subsec:purity_completeness}), we can write the expected group catalogue inferred red satellite fraction for the entire sample as,
\begin{align}
f_{\rm sat|red}^{\rm inf} = & [ C_{\rm sat}f_{\rm red|sat}^{\rm mock}N_{\rm sat}^{\rm mock} \\ \nonumber
&+(1-C_{\rm cen})f_{\rm red|cen}^{\rm mock}N_{\rm cen}^{\rm mock}] / (f_{\rm red} N_{\rm gal}),
\end{align}
where, for example, $f_{\rm red|sat}^{\rm mock}$ is the fraction of satellites in the mock which are red.  However, using this calculation, one would expect the {\tt TINKER-GF} groups inferred red (blue) satellite fraction to be under-estimated (over-estimated) by $\sim 11 \%$ ($29  \%$); however, empirically, it is under-estimated (over-estimated) by only $\sim 4 \%$ ($12\%$) for the full sample.  For a mock with weaker 2-halo conformity, the colour-dependent satellite fractions will have a larger error.

\subsubsection{Galactic Conformity}
\label{subsec:discussion_conformity}

A primary motivation for our study of group finding errors was to investigate the ability of galaxy group catalogues to study the galactic conformity phenomenon.  In \S~\ref{subsec:results_conformity}, we measured the blue fraction of satellites around blue and red centrals as a function of halo mass.  This is consistent with the original measurement of conformity by \cite{Weinmann:2006hu}, who find that at fixed halo mass, groups with a passive central galaxy have, on average, a more passive satellite population than groups with a star forming central galaxy.  We find that for our mock, one with strong conformity of the magnitude measured by \cite{Weinmann:2006hu}, group finders largely recover the 1-halo conformity signal.  However, we also found that group finders tend to induce a small conformity signal in a mock constructed to have no conformity.  The primary difficulties in measuring 1-halo conformity are twofold: accounting for systematic halo mass differences in groups with red and blue centrals, and the effect of membership errors in groups that results from fusing and fracturing.  As we have seen, both of these can induce a false conformity-like signal. 

A subsequent measurement of 1-halo conformity using the Yang et al group catalogue was made by \cite{Knobel:2014tx}, who further control for additional environmental variables beyond halo mass, finding a result in qualitative agreement with \cite{Weinmann:2006hu}, namely that satellites around quenched centrals display an increased quenching efficiency parameter compared to those around star forming centrals.  \cite{Knobel:2014tx} attribute the persistence of the galactic conformity signal to ``hidden variables'' not controlled for in the sample selection when comparing satellite samples, including uncertainties in the estimation of halo mass, or physical parameters like halo formation time \citep[i.e. $z_{\rm form}$ in][]{Hearin:2013km}.  Differentiating between physical effects and systematics in measuring group properties remains a challenge.     

The problems involved in using galaxy group catalogues to measure 1-halo conformity extend to studies which use alternate methods.  \cite{Phillips:2014bt,Phillips:2014vt} instead use an isolation criteria, in essence a type of conservative group finder, to select $L^{*}$ central-$0.1 L^{*}$ satellite spectroscopic pairs in the SDSS.  With this sample, they study the statistical properties of the satellite populations around star forming and quiescent $L^{*}$ galaxies.  They find that star forming central galaxies have satellite populations that are more star forming than quiescent centrals at a fixed central stellar mass.  They point out that their measurement of conformity could be explained if quenching efficiency has a strong dependence on halo mass, especially for haloes $\sim 10^{12}M_{\odot}$.  Beyond halo mass estimation difficulties, \cite{Phillips:2014vt} use a correction scheme based on purity and completeness, similar to the $P$ and $C$ measures discussed in \S~\ref{subsec:purity_completeness}, to account for interlopers in their satellite sample.  This correction may over or under-correct the inferred statistics, or be more or less applicable to the sample split into star forming and passive sub-samples.  

The primary challenge for conformity measurements is to isolate samples of star forming and quiescent galaxies at fixed halo mass.  Group catalogues provide one method to attempt this, along with isolation criteria, but they are subject to systematic effects of the type discussed in this paper.  Satellite velocity distributions and lensing measurements around groups or `isolated' galaxies may provide one mechanism to reduce systematic uncertainties in future measurements of conformity.

\subsection{The Role of Halo Definitions}
\label{subsec:discussion_halodef}

The group finding process is sensitive to the precise halo definition adopted.  In a simulation, the boundary of a halo will define which dark matter structures are subhaloes (those within the boundary), and which are neighbouring haloes (those outside the boundary).  If a group finder is optimized to recover a galaxy population residing in a halo of a particular definition, and the model that the results are being compared to defines haloes in a different manner, the results are difficult to interpret.  On the other hand, this problem gets to the root of the challenge facing any group finder, namely optimizing the group finder to recover physically meaningful groups while minimizing membership errors.

Each group finder discussed in this paper was tuned on a different set of mocks with a different set of criteria setting the tuning parameters.  In particular, the {\tt YANG-GF} and {\tt TINKER-GF} were tuned on SO halo based mocks to reproduce SO halo galaxy populations but with different optimization criteria.  The former minimizes interlopers and maximizes completeness on the group to group level while the latter is optimized to reproduce the full HOD of the mock.  The {\tt BERLIND-GF} was tuned on FoF halo based mocks, with a different set of optimization goals geared towards reproducing the FoF halo population characteristics of higher mass groups than the {\tt YANG-GF} and {\tt TINKER-GF}.

One result of these different halo definitions and optimization schemes is apparent throughout much of this paper.  The {\tt YANG-GF} and {\tt TINKER-GF} groups are often in closer agreement to the mock values (w/ SO haloes), especially for measurements of occupation statistics sensitive to the over-all number of galaxies in a halo such as the $\langle N_{\rm gal}|M_{\rm vir} \rangle$ and $f_{\rm sat}$.  This is because the {\tt BERLIND-GF} groups are geared to reproduce FoF haloes which trace different structures than SO haloes.  Additionally, the {\tt BERLIND-GF} is optimized to reproduce relatively large groups compared to the other two group finders. The result is that the {\tt BERLIND-GF} group catalogue has an elevated satellite fraction of $\sim45\%$ relative to the mocks intrinsic value of  $\sim25\%$. 

While the number of galaxies in haloes is sensitive to the halo definition used,  the relative abundance of red and blue galaxies need not to be affected in the same way.  The age-matching mock used for this study assigns galaxy colour based on a measure of halo age, $z_{\rm form}$, and the age parameter of satellites is very rarely affected by their accretion into a host halo.  The result is that there is no abrupt colour transition between satellites and galaxies outside their halo, with weak colour gradients on the 1-halo to 2-halo scale.  This is a manifestation of the 2-halo galactic conformity phenomenon.  As discussed throughout the paper, it is this 2-halo conformity phenomenon which results in smaller errors in the group finder inferred colour dependent occupation statistics as a result of interlopers than one would expect if there were no 2-halo colour correlations.  A corollary to this is that small changes to the boundary of a halo should not drastically affect relative abundance of red and blue satellites in groups, i.e. measurements $f_{\rm red}$.                   

We do not attempt to retune any of the group finders used in this paper for a few reasons.  First, it is not possible to retune a group finder on the real universe, so not retuning on our particular mock gives a more useful picture of the kind and magnitude of the errors group finders are likely to make.  Second, it is interesting to note the effect of the different optimization goals and halo definitions on the outcome of the colour-dependent statistics measured here.  Furthermore, we use the group finders as they are because they have often been used in the literature with the tunings used here, and we believe it is important to show the results as they are with no further optimization.  It is an interesting question to ask which group finding method and set of tuning parameters is best suited to recover a particular occupation statistic, but such a study is beyond the scope of this work.

\subsection{Additional Systematics} 
\label{subsec:primary_prop}
Our mock and group finders assume that galaxy luminosity is the primary galaxy property which determines the galaxy occupation statistics.  This is a common assumption, but another galaxy property, namely stellar mass, could instead have been assumed to be the primary property.  We have tested that the qualitative features of the mock remain unchanged if instead we use stellar mass as the primary galaxy property \citep[see][]{Hearin:2014hh}.  Using a `perfect' group finder analysis similar to the method discussed in \S~\ref{subsec:results_group_vs_halo}, we have checked that the three error modes described in this paper remain unchanged in a stellar mass based mock.  As the primary goal of this paper is not to re-tune group finders as discussed previously, we do not retune each one to use stellar mass as the halo mass indicator \citep[see][for a discussion of this effect]{Yang:2007db}.   

\cite{Yang:2008tg} examine the affect of using stellar mass and luminosity as the primary galaxy property on inferring the halo mass dependence of the central galaxy quenched fraction and the colour split satellite fraction (see Fig. 5 and 11 therein respectively).  The difference between the two measures can be significant when the statistic is dependent on the halo mass estimation, indicating that this can be a significant and even dominant source of uncertainty.  In a forward modelling approach, the trends that one is able to infer from a group catalogue will only be as accurate as the mock and the power of the observational statistic.  The ability to distinguish between galaxy luminosity and stellar mass as the primary galaxy property driving halo occupation requires further work.  

\begin{figure}
	\includegraphics[]{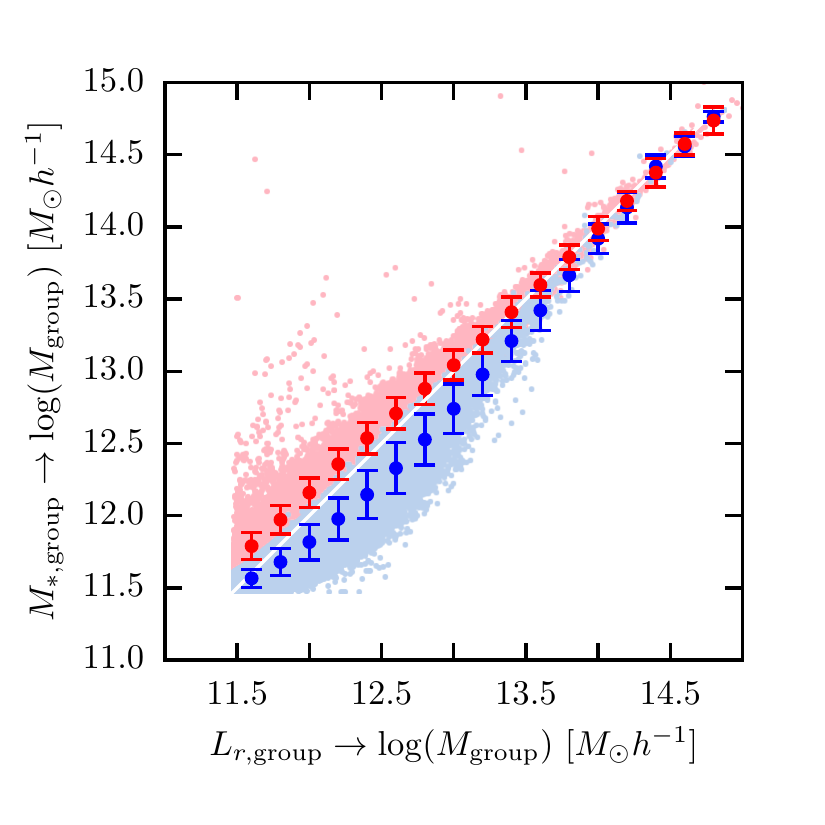} 
	\caption{inferred halo mass using total group stellar mass vs. inferred halo mass using total group luminosity for groups with red central galaxies (red points with error bars indicating the scatter) and blue central galaxies (blue points with error bars indicating the scatter).  When galaxy luminosity is assumed to be the primary galaxy property determining halo occupation to build a mock, and group halo mass is estimated with total stellar mass, groups with red galaxies are systematically assigned more massive halo mass estimates.  Extreme outliers are a result of galaxies with extremely red colours and the overly simple mass-to-light ratio conversion used to calculate stellar mass.}
\label{fig:halo_mass_systematics}
\end{figure}

Never-the-less, we briefly discuss the affect of using the `incorrect' property on our group analysis, in the case of our mocks, stellar mass.  That is, what is the effect of assuming stellar mass as the primary halo occupation indicator for an analysis of group finders run over our mock where luminosity was assumed to be the primary indicator?  We examined the affect using a `perfect' group finder analysis, assigning every galaxy a stellar mass using the relation between mass-to-light and colour from \cite{Bell:2003hs} and used by \cite{Yang:2007db} where
\begin{align}
\log(M_*/(h^{-2}M_{\odot}))=&-0.406 +1.097\left[{}^{0.0}(g-r)\right] \nonumber \\
                                             &-0.4({}^{0.0}M_r-5\log(h)-4.64)
\end{align}
where the $0.0$ indicates the k-corrected quantity to $z=0.0$. We then use total group stellar mass to estimate the halo mass of groups.  The primary error induced by this assumption is that redder galaxies are inferred to occupy more massive haloes at fixed luminosity, the effect being strongest in groups with fewer members.  To see the magnitude of this error, in Fig. \ref{fig:halo_mass_systematics} we show the comparison between group masses calculated both ways.  The systematic mass error between red and blue galaxies can be of the order of 0.5 dex.

One interesting result of this concerns galactic conformity, where the mass bias effect is strong enough to infer a significantly reduced galactic conformity signal in our mock with intrinsic conformity.  No conformity signal is induced in the mock without intrinsic conformity.  This occurs because groups with red centrals are inferred to be more massive than their {\em true} mass, bringing their intrinsically higher red fraction of satellites more in line with groups of the true inferred mass.  This is contrary to the intrinsic conformity signal in a version of our mock built using stellar mass and sSFR.  This result is also contrary to an empirical analysis of SDSS group catalogues which continue to display 1-halo galactic conformity when using stellar mass as the halo mass indicator \citep[see][]{Kauffmann:2013jd, Knobel:2014tx,Phillips:2014bt}.  We leave an exhaustive examination of 1-halo conformity in SDSS to a forthcoming paper.      
  
It is also important to note that we have also neglected to include a model for spectroscopic incompleteness in our study.  The most significant source of incompleteness in the SDSS is due to the fact that two fibers used to obtain spectra cannot be placed within $55''$.  Approximately $\sim7\%$ of galaxies are affected in the SDSS main sample, disproportionately affecting galaxies in regions of high-density.  Those that redshifts are not obtained for are called fiber collisions.  These galaxies are often included in the spectroscopic galaxy sample used to build group catalogues by using the redshift of the nearest neighbour.  For roughly $60\%$ of galaxies, this is a good approximation \citep{Zehavi:2002dd}.  The remainder of the time, this assigns a redshift to a galaxy that is significantly in error.  Including these galaxies in an analysis will generally decrease the purity of group memberships, and excluding them will decrease the completeness. Furthermore, because fiber collisions preferentially occur in dense regions, and galaxy colour is correlated with density, including or excluding these galaxies will likely result in some un-modelled correlated colour error.  This affect is easily studied empirically in an analysis by including and excluding affected galaxies and examining the result on a statistic.   We do not address this source of error in our analysis, and refer the reader to \cite{Yang:2007db} for a more lengthy discussion on the affect of fiber collisions on the Yang et al. group finder, where indications are that the effects are small.

\subsection{Improving Galaxy Group Catalogues}
\label{subsec:discussion_correcting_gf}

There have been several attempts to ``correct'' galaxy group catalogue statistics using the purity and completeness measures discussed in \S~\ref{subsec:purity_completeness} and appendix \ref{appendix:group_finding_errors}.  \citet{Tinker:2011ek} empirically measured the purity and completeness of satellites in groups as a function of large scale environment.  They then use this average measurement to correct their results assuming interlopers are taken at random from the central population.  As previously noted, \cite{Phillips:2014bt,Phillips:2014vt} preform a very similar operation on their measurements of central-satellite pairs. 

In principle, the HTP (\S~\ref{subsec:HTP}) contains all the information necessary to correct 1-point statistics.  The feasibility and accuracy of such a scheme is dependent on the stability of the HTP to the underlying models which one seeks to differentiate between.  We leave such a study to future work, but briefly describe what would be required.  First, a large number of mocks with  consistent clustering statistics, but varying models for assigning the relevant galaxy property, e.g. colour, would need to be created.  A group finder would then need to be run over each mock, calculating the HTP for each.  The variation in the ensemble of HTPs could then be quantified, and if the variation is low, then, in principle, the true underlying colour-dependent statistics can be measured utilizing the stable HTP. 

Beyond utilization of the HTP, group finders may be improved in other ways.  We use a simple prescription to distinguish a central galaxy from satellites in a group, namely we assign the most luminous galaxy in the group to be the central galaxy.  We expect this to fail some fraction of the time, as demonstrated in \S~\ref{subsec:cen_sat_errors}, even if group finders assign galaxy membership perfectly, and we see evidence for this failure in the HTPs.  It is possible that by using geometric and velocity information from high multiplicity groups that central assignment accuracy could be increased.  Improving this for lower mass groups with fewer members remains difficult.  We did not investigate this issue here and leave it to the consideration of future work.

The group finding process could be improved and tuned for specific tasks.  \cite{Duarte:2014ug} analysed the failure modes of FoF based grouping algorithms and quantified the regimes where different linking lengths are most successful in recovering group properties.  This suggests an adaptive FoF linking algorithm may be best suited for recovering groups over a wide range of environments and masses, an approach which begins to look more similar to the {\tt YANG-GF} and {\tt TINKER-GF} methods.    

Ideally, group finders should be used to build groups in a probabilistic sense to capture the inherent uncertainties in the process \citep[e.g. in the vain of][]{Botzler:2004hf, Li:2008kw, Jian:2014rv,Duarte:2014ue}.  Given some model, it is possible to assign probabilities that individual galaxies reside in a common group--currently this information is explicitly ignored in \citep{Yang:2005cn,Yang:2007db}.  During the final stages of preparation of this manuscript \cite{Duarte:2014ue} have explicitly retained and used probabilities of group membership in the formulation of a new group finder, MAGGIE.  While it would be very interesting to investigate whether MAGGIE, or another probabilistic approach to the the grouping problem, alleviates some of the issues investigated here, we leave this to future work.

Finally, statistics should not be directly inferred from galaxy group catalogues without careful thought as to how the errors explored in this paper could affect the measurements.  The most straightforward solution is to forward model the group finding process before making comparisons to observational data \cite[see][for an example]{Wetzel:2013dw}.  This can sometimes require the creation of novel mock galaxy catalogues, but this approach can be immediately fruitful without modification to current group finding algorithms.

\section{Summary}
\label{sec:summary}

We described three types of errors group finding is subject to.  These are errors in:
\begin{enumerate}
\item halo mass estimation
\item central/satellite designation
\item group membership determination
\end{enumerate}
and in practice, errors in each of these categories occur and are coupled.  When galaxies that reside in a common halo are misidentified to be in two or more groups, a process called fracturing, group finders tend to misidentify satellites as centrals and underestimate group mass.  When galaxies that reside in two or more haloes are misidentified to be in a common group, a process called fusing, group finders tend to misidentify centrals as satellites and overestimate group mass.  For a given group finder (and mock), this correlated error is characterized by the halo transition probability (HTP).                

Errors in group finding will affect inferred colour-dependent occupation statistics measured directly with a galaxy group catalogue.  To characterize the specific errors made, we used three group finders to measure colour-dependent occupation statistics on a mock galaxy survey created using the age-matching technique.  For the set of occupation statistics explored in this study, group finder inferred measurements:  
\begin{enumerate}
\item tend to equalize the properties of distinct galaxy sub-populations
\item are able to recover 1-halo conformity (but also induce a small 1-halo conformity signal)
\item errors are reduced by the presence of realistic 2-halo conformity in our mock.
\end{enumerate}

We examined the ability to directly infer colour dependent occupation statistics $\langle N_{\rm gal}|M\rangle$, $f_{\rm sat}(L_{\rm gal})$, $\phi(L_{\rm gal}|M){\rm d}L$, $f_{\rm red}(M_{\rm halo})$, and 1-halo conformity from galaxy group catalogues.  The first moment of the HOD is largely recovered for the full sample of galaxies, only varying significantly from the true relation for the sub-dominant blue galaxy sub-population.  The satellite fraction as a function of galaxy luminosity is sensitive to the halo definition applied when tuning a group finder, but when this definition is close to that of the underlying mock, group finders recover this statistics very well.  The CLF is also recovered accurately for the full galaxy sample, only showing some systematic deviation for the colour split sample where the sub-dominant blue population of centrals and satellites is over estimated.  The red fraction of galaxies as a function of halo mass is poorly recovered by group finders, particularly the central red fraction, an important result that may have implications for the inference halo-galaxy quenching relations.  Group finders recover a 1-halo conformity signal of the same magnitude as the underlying signal when one exists, but display a tendency to induce a signal in a mock constructed to have no conformity.     

To conclude, colour-dependent occupation statistics inferred from galaxy group catalogues are affected by group finding errors in non-trivial ways.  Measurements of galaxy(-group) statistics requires fully modelling the group finding process and its errors  using the HTP, or by forward modelling the group finding process to make a fair comparison to data.

\section*{Acknowledgements}
DC would like to thank Erik Tollerud for helpful discussions and B.S. for a forum for discussions that added to this paper.  APH thanks Doug Watson for insightful jam sessions.

\bibliographystyle{mn2elong}
\bibliography{bib}

\begin{thebibliography}{112}
\expandafter\ifx\csname natexlab\endcsname\relax\def\natexlab#1{#1}\fi

\bibitem[{Balogh {et~al}\mbox{.}(2004)Balogh, Baldry, Nichol, Miller, Bower, \&
  Glazebrook}]{Balogh:2004cx}
Balogh M.~L., Baldry I.~K., Nichol R., Miller C., Bower R., Glazebrook K.,
  2004, ApJ, 615, L101

\bibitem[{Balogh {et~al}\mbox{.}(2000)Balogh, Navarro, \&
  Morris}]{Balogh:2000hn}
Balogh M.~L., Navarro J.~F., Morris S.~L., 2000, ApJ, 540, 113

\bibitem[{Behroozi {et~al}\mbox{.}(2013)Behroozi, Wechsler, \&
  Wu}]{Behroozi:2013cn}
Behroozi P.~S., Wechsler R.~H., Wu H.-Y., 2013, ApJ, 762, 109

\bibitem[{Behroozi {et~al}\mbox{.}(2012)Behroozi, Wechsler, Wu, Busha, Klypin,
  \& Primack}]{Behroozi:2012dz}
Behroozi P.~S., Wechsler R.~H., Wu H.-Y., Busha M.~T., Klypin A.~A., Primack
  J.~R., 2012, ApJ, 763, 18

\bibitem[{Bell {et~al}\mbox{.}(2003)Bell, McIntosh, Katz, \&
  Weinberg}]{Bell:2003hs}
Bell E.~F., McIntosh D.~H., Katz N., Weinberg M.~D., 2003, ApJS, 149, 289

\bibitem[{Bell {et~al}\mbox{.}(2004)Bell, Wolf, Meisenheimer, Rix, Borch, Dye,
  Kleinheinrich, Wisotzki, \& McIntosh}]{Bell:2004fo}
Bell E.~F. {et~al.}, 2004, ApJ, 608, 752

\bibitem[{Berlind {et~al}\mbox{.}(2006)Berlind, Frieman, Weinberg, Blanton,
  Warren, Abazajian, Scranton, Hogg, Scoccimarro, Bahcall, Brinkmann, Gott,
  Kleinman, Krzesinski, Lee, Miller, Nitta, Schneider, Tucker, Zehavi, \&
  Collaboration}]{Berlind:2006dy}
Berlind A.~A. {et~al.}, 2006, ApJS, 167, 1

\bibitem[{Blanton {et~al}\mbox{.}(2003{\natexlab{a}})Blanton, Brinkmann,
  Csabai, Doi, Eisenstein, Fukugita, Gunn, Hogg, \& Schlegel}]{Blanton:2003iu}
Blanton M.~R. {et~al.}, 2003{\natexlab{a}}, AJ, 125, 2348

\bibitem[{Blanton {et~al}\mbox{.}(2005{\natexlab{a}})Blanton, Eisenstein, Hogg,
  Schlegel, \& Brinkmann}]{Blanton:2005eb}
Blanton M.~R., Eisenstein D., Hogg D.~W., Schlegel D.~J., Brinkmann J.,
  2005{\natexlab{a}}, ApJ, 629, 143

\bibitem[{Blanton {et~al}\mbox{.}(2003{\natexlab{b}})Blanton, Hogg, Bahcall,
  Baldry, Brinkmann, Csabai, Eisenstein, Fukugita, Gunn, Ivezi{\'c}, Lamb,
  Lupton, Loveday, Munn, Nichol, Okamura, Schlegel, Shimasaku, Strauss,
  Vogeley, \& Weinberg}]{Blanton:2003hh}
Blanton M.~R. {et~al.}, 2003{\natexlab{b}}, ApJ, 594, 186

\bibitem[{Blanton {et~al}\mbox{.}(2005{\natexlab{b}})Blanton, Schlegel,
  Strauss, Brinkmann, Finkbeiner, Fukugita, Gunn, Hogg, Ivezi{\'c}, Knapp,
  Lupton, Munn, Schneider, Tegmark, \& Zehavi}]{Blanton:2005fs}
Blanton M.~R. {et~al.}, 2005{\natexlab{b}}, AJ, 129, 2562

\bibitem[{Botzler {et~al}\mbox{.}(2004)Botzler, Snigula, Bender, \&
  Hopp}]{Botzler:2004hf}
Botzler C.~S., Snigula J., Bender R., Hopp U., 2004, MNRAS, 349, 425

\bibitem[{Cacciato {et~al}\mbox{.}(2013)Cacciato, van~den Bosch, More, Mo, \&
  Yang}]{Cacciato:2013dd}
Cacciato M., van~den Bosch F.~C., More S., Mo H., Yang X., 2013, MNRAS, 430,
  767

\bibitem[{Calvi {et~al}\mbox{.}(2011)Calvi, Poggianti, \&
  Vulcani}]{Calvi:2011hp}
Calvi R., Poggianti B.~M., Vulcani B., 2011, MNRAS, 416, 727

\bibitem[{Conroy {et~al}\mbox{.}(2006)Conroy, Wechsler, \&
  Kravtsov}]{Conroy:2006iz}
Conroy C., Wechsler R.~H., Kravtsov A.~V., 2006, ApJ, 647, 201

\bibitem[{Cooper {et~al}\mbox{.}(2007)Cooper, Newman, Coil, Croton, Gerke, Yan,
  Davis, Faber, Guhathakurta, Koo, Weiner, \& Willmer}]{Cooper:2007gn}
Cooper M.~C. {et~al.}, 2007, MNRAS, 376, 1445

\bibitem[{Cooray(2006)}]{Cooray:2006ek}
Cooray A., 2006, MNRAS, 365, 842

\bibitem[{Cooray \& Milosavljevi{\'c}(2005)}]{Cooray:2005gd}
Cooray A., Milosavljevi{\'c} M., 2005, ApJ, 627, L85

\bibitem[{Cucciati {et~al}\mbox{.}(2006)Cucciati, Iovino, Marinoni, Ilbert,
  Bardelli, Franzetti, Le~F{\`e}vre, Pollo, Zamorani, Cappi, Guzzo, McCracken,
  Meneux, Scaramella, Scodeggio, Tresse, Zucca, Bottini, Garilli, Le~Brun,
  Maccagni, Picat, Vettolani, Zanichelli, Adami, Arnaboldi, Arnouts,
  Bolzonella, Charlot, Ciliegi, Contini, Foucaud, Gavignaud, Marano, Mazure,
  Merighi, Paltani, Pell{\`o}, Pozzetti, Radovich, Bondi, Bongiorno, Busarello,
  de~la Torre, Gregorini, Lamareille, Mathez, Mellier, Merluzzi, Ripepi, Rizzo,
  Temporin, \& Vergani}]{Cucciati:2006fp}
Cucciati O. {et~al.}, 2006, A{\&}A, 458, 39

\bibitem[{Dressler(1980)}]{Dressler:1980ie}
Dressler A., 1980, ApJ, 236, 351

\bibitem[{Duarte \& Mamon(2014{\natexlab{a}})}]{Duarte:2014ug}
Duarte M., Mamon G.~A., 2014{\natexlab{a}}, MNRAS, 440, 1763

\bibitem[{Duarte \& Mamon(2014{\natexlab{b}})}]{Duarte:2014ue}
Duarte M., Mamon G.~A., 2014{\natexlab{b}}, \eprint{1412.3364}

\bibitem[{Dubinski(1998)}]{Dubinski:1998bz}
Dubinski J., 1998, ApJ, 502, 141

\bibitem[{Eke {et~al}\mbox{.}(2004)Eke, Baugh, Cole, Frenk, Norberg, Peacock,
  Baldry, Bland-Hawthorn, Bridges, Cannon, Colless, Collins, Couch, Dalton,
  de~Propris, Driver, Efstathiou, Ellis, Glazebrook, Jackson, Lahav, Lewis,
  Lumsden, Maddox, Madgwick, Peterson, Sutherland, \& Taylor}]{Eke:2004bf}
Eke V.~R. {et~al.}, 2004, MNRAS, 348, 866

\bibitem[{Farouki \& Shapiro(1981)}]{Farouki:1981br}
Farouki R., Shapiro S.~L., 1981, ApJ, 243, 32

\bibitem[{Geller \& Huchra(1983)}]{Geller:1983ih}
Geller M.~J., Huchra J.~P., 1983, ApJS, 52, 61

\bibitem[{Giuricin {et~al}\mbox{.}(2000)Giuricin, Marinoni, Ceriani, \&
  Pisani}]{Giuricin:2000eg}
Giuricin G., Marinoni C., Ceriani L., Pisani A., 2000, ApJ, 543, 178

\bibitem[{Grebel {et~al}\mbox{.}(2003)Grebel, Gallagher, \&
  Harbeck}]{Grebel:2003iw}
Grebel E.~K., Gallagher J. S.~I., Harbeck D., 2003, AJ, 125, 1926

\bibitem[{Hansen {et~al}\mbox{.}(2009)Hansen, Sheldon, Wechsler, \&
  Koester}]{Hansen:2009if}
Hansen S.~M., Sheldon E.~S., Wechsler R.~H., Koester B.~P., 2009, ApJ, 699,
  1333

\bibitem[{Hartley {et~al}\mbox{.}(2014)Hartley, Conselice, Mortlock, Foucaud,
  \& Simpson}]{Hartley:2014uq}
Hartley W.~G., Conselice C.~J., Mortlock A., Foucaud S., Simpson C., 2014,
  \eprint{1406.6058v1}

\bibitem[{Hearin \& Watson(2013)}]{Hearin:2013km}
Hearin A.~P., Watson D.~F., 2013, MNRAS, 435, 1313

\bibitem[{Hearin {et~al}\mbox{.}(2014{\natexlab{a}})Hearin, Watson, Becker,
  Reyes, Berlind, \& Zentner}]{Hearin:2014hh}
Hearin A.~P., Watson D.~F., Becker M.~R., Reyes R., Berlind A.~A., Zentner
  A.~R., 2014{\natexlab{a}}, MNRAS, 444, 729

\bibitem[{Hearin {et~al}\mbox{.}(2014{\natexlab{b}})Hearin, Watson, \& van~den
  Bosch}]{Hearin:2014wv}
Hearin A.~P., Watson D.~F., van~den Bosch F.~C., 2014{\natexlab{b}},
  \eprint{1404.6524}

\bibitem[{Hearin {et~al}\mbox{.}(2013)Hearin, Zentner, Berlind, \&
  Newman}]{Hearin:2013ok}
Hearin A.~P., Zentner A.~R., Berlind A.~A., Newman J.~A., 2013, MNRAS, 433, 659

\bibitem[{Hogg {et~al}\mbox{.}(2004)Hogg, Blanton, Brinchmann, Eisenstein,
  Schlegel, Gunn, McKay, Rix, Bahcall, Brinkmann, \& Meiksin}]{Hogg:2004ev}
Hogg D.~W. {et~al.}, 2004, ApJ, 601, L29

\bibitem[{Hou {et~al}\mbox{.}(2014)Hou, Parker, \& Harris}]{Hou:2014yg}
Hou A., Parker L.~C., Harris W.~E., 2014, MNRAS, 442, 406

\bibitem[{Huchra \& Geller(1982)}]{Huchra:1982ed}
Huchra J.~P., Geller M.~J., 1982, ApJ, 257, 423

\bibitem[{Jian {et~al}\mbox{.}(2014)Jian, Lin, Chiueh, Lin, Liu, Merson, Baugh,
  Huang, Chen, Foucaud, Murphy, Cole, Burgett, \& Kaiser}]{Jian:2014rv}
Jian H.-Y. {et~al.}, 2014, ApJ, 788, 109

\bibitem[{Kauffmann {et~al}\mbox{.}(2013)Kauffmann, Kauffmann, Li, Li, Zhang,
  Zhang, Weinmann, \& Weinmann}]{Kauffmann:2013jd}
Kauffmann G., Kauffmann G., Li C., Li C., Zhang W., Zhang W., Weinmann S.,
  Weinmann S., 2013, MNRAS, 430, 1447

\bibitem[{Kauffmann {et~al}\mbox{.}(2004)Kauffmann, White, Heckman, M{\'e}nard,
  Brinchmann, Charlot, Tremonti, \& Brinkmann}]{Kauffmann:2004cw}
Kauffmann G., White S. D.~M., Heckman T.~M., M{\'e}nard B., Brinchmann J.,
  Charlot S., Tremonti C., Brinkmann J., 2004, MNRAS, 353, 713

\bibitem[{Kawata \& Mulchaey(2008)}]{Kawata:2008cd}
Kawata D., Mulchaey J.~S., 2008, ApJ, 672, L103

\bibitem[{Klypin {et~al}\mbox{.}(2011)Klypin, Trujillo-Gomez, \&
  Primack}]{Klypin:2011bd}
Klypin A.~A., Trujillo-Gomez S., Primack J., 2011, ApJ, 740, 102

\bibitem[{Knebe {et~al}\mbox{.}(2011)Knebe, Knollmann, Muldrew, Pearce,
  Aragon-Calvo, Ascasibar, Behroozi, Ceverino, Colombi, Diemand, Dolag, Falck,
  Fasel, Gardner, Gottl{\"o}ber, Hsu, Iannuzzi, Klypin, Luki{\'c}, Maciejewski,
  McBride, Neyrinck, Planelles, Potter, Quilis, Rasera, Read, Ricker, Roy,
  Springel, Stadel, Stinson, Sutter, Turchaninov, Tweed, Yepes, \&
  Zemp}]{Knebe:2011jc}
Knebe A. {et~al.}, 2011, MNRAS, 415, 2293

\bibitem[{Knebe {et~al}\mbox{.}(2013)Knebe, Pearce, Lux, Ascasibar, Behroozi,
  Casado, Moran, Diemand, Dolag, Dominguez-Tenreiro, Elahi, Falck,
  Gottl{\"o}ber, Han, Klypin, Luki{\'c}, Maciejewski, McBride, Merchan,
  Muldrew, Neyrinck, Onions, Planelles, Potter, Quilis, Rasera, Ricker, Roy,
  Ruiz, Sgro, Springel, Stadel, Sutter, Tweed, \& Zemp}]{Knebe:2013bp}
Knebe A. {et~al.}, 2013, MNRAS, 435, 1618

\bibitem[{Knobel {et~al}\mbox{.}(2014)Knobel, Lilly, Woo, \& Kova{\v
  c}}]{Knobel:2014tx}
Knobel C., Lilly S.~J., Woo J., Kova{\v c} K., 2014, \eprint{1408.2553}

\bibitem[{Kravtsov {et~al}\mbox{.}(2004)Kravtsov, Berlind, Wechsler, Klypin,
  Gottl{\"o}ber, Allgood, \& Primack}]{Kravtsov:2004fi}
Kravtsov A.~V., Berlind A.~A., Wechsler R.~H., Klypin A.~A., Gottl{\"o}ber S.,
  Allgood B., Primack J.~R., 2004, ApJ, 609, 35

\bibitem[{Kravtsov {et~al}\mbox{.}(1997)Kravtsov, Klypin, \&
  Khokhlov}]{Kravtsov:1997iy}
Kravtsov A.~V., Klypin A.~A., Khokhlov A.~M., 1997, ApJS, 111, 73

\bibitem[{Larson {et~al}\mbox{.}(1980)Larson, Tinsley, \&
  Caldwell}]{Larson:1980id}
Larson R.~B., Tinsley B.~M., Caldwell C.~N., 1980, ApJ, 237, 692

\bibitem[{Li \& Yee(2008)}]{Li:2008kw}
Li I.~H., Yee H. K.~C., 2008, AJ, 135, 809

\bibitem[{McCarthy {et~al}\mbox{.}(2008)McCarthy, Frenk, Font, Lacey, Bower,
  Mitchell, Balogh, \& Theuns}]{McCarthy:2008en}
McCarthy I.~G., Frenk C.~S., Font A.~S., Lacey C.~G., Bower R.~G., Mitchell
  N.~L., Balogh M.~L., Theuns T., 2008, MNRAS, 383, 593

\bibitem[{Merch{\'a}n \& Zandivarez(2002)}]{Merchan:2002gp}
Merch{\'a}n M., Zandivarez A., 2002, MNRAS, 335, 216

\bibitem[{Merchan \& Zandivarez(2005)}]{Merchan:2005gc}
Merchan M.~E., Zandivarez A., 2005, ApJ, 630, 759

\bibitem[{Miller {et~al}\mbox{.}(2005)Miller, Nichol, Reichart, Wechsler,
  Evrard, Annis, McKay, Bahcall, Bernardi, Boehringer, Connolly, Goto, Kniazev,
  Lamb, Postman, Schneider, Sheth, \& Voges}]{Miller:2005kc}
Miller C.~J. {et~al.}, 2005, AJ, 130, 968

\bibitem[{Moore {et~al}\mbox{.}(1993)Moore, Frenk, \& White}]{Moore:1993kv}
Moore B., Frenk C.~S., White S. D.~M., 1993, MNRAS, 261, 827

\bibitem[{Moore {et~al}\mbox{.}(1998)Moore, Lake, \& Katz}]{Moore:1998jz}
Moore B., Lake G., Katz N., 1998, ApJ, 495, 139

\bibitem[{More {et~al}\mbox{.}(2011)More, van~den Bosch, Cacciato, Skibba, Mo,
  \& Yang}]{More:2011il}
More S., van~den Bosch F.~C., Cacciato M., Skibba R., Mo H.~J., Yang X., 2011,
  MNRAS, 410, 210

\bibitem[{Mu{\~n}oz-Cuartas \& M{\"u}ller(2012)}]{MunozCuartas:2012cx}
Mu{\~n}oz-Cuartas J.~C., M{\"u}ller V., 2012, MNRAS, 423, 1583

\bibitem[{Navarro {et~al}\mbox{.}(1997)Navarro, Frenk, \&
  White}]{Navarro:1997if}
Navarro J.~F., Frenk C.~S., White S. D.~M., 1997, 490, 493

\bibitem[{Nolthenius \& White(1987)}]{Nolthenius:1987vf}
Nolthenius R., White S. D.~M., 1987, MNRAS, 225, 505

\bibitem[{Pasquali {et~al}\mbox{.}(2010)Pasquali, Gallazzi, Fontanot, van~den
  Bosch, De~Lucia, Mo, \& Yang}]{Pasquali:2010ky}
Pasquali A., Gallazzi A., Fontanot F., van~den Bosch F.~C., De~Lucia G., Mo
  H.~J., Yang X., 2010, MNRAS, 407, 937

\bibitem[{Pasquali {et~al}\mbox{.}(2009)Pasquali, van~den Bosch, Mo, Yang, \&
  Somerville}]{Pasquali:2009fi}
Pasquali A., van~den Bosch F.~C., Mo H.~J., Yang X., Somerville R., 2009,
  MNRAS, 394, 38

\bibitem[{Peng {et~al}\mbox{.}(2010)Peng, Lilly, Kova{\v c}, Bolzonella,
  Pozzetti, Renzini, Zamorani, Ilbert, Knobel, Iovino, Maier, Cucciati, Tasca,
  Carollo, Silverman, Kampczyk, de~Ravel, Sanders, Scoville, Contini, Mainieri,
  Scodeggio, Kneib, Le~F{\`e}vre, Bardelli, Bongiorno, Caputi, Coppa, de~la
  Torre, Franzetti, Garilli, Lamareille, Le~Borgne, Le~Brun, Mignoli,
  Perez~Montero, Pello, Ricciardelli, Tanaka, Tresse, Vergani, Welikala, Zucca,
  Oesch, Abbas, Barnes, Bordoloi, Bottini, Cappi, Cassata, Cimatti, Fumana,
  Hasinger, Koekemoer, Leauthaud, Maccagni, Marinoni, McCracken, Memeo, Meneux,
  Nair, Porciani, Presotto, \& Scaramella}]{Peng:2010gn}
Peng Y.-j. {et~al.}, 2010, ApJ, 721, 193

\bibitem[{Peng {et~al}\mbox{.}(2012)Peng, Lilly, Renzini, \&
  Carollo}]{Peng:2012bp}
Peng Y.-j., Lilly S.~J., Renzini A., Carollo M., 2012, ApJ, 757, 4

\bibitem[{Phillips {et~al}\mbox{.}(2014{\natexlab{a}})Phillips, Wheeler,
  Boylan-Kolchin, Bullock, Cooper, \& Tollerud}]{Phillips:2014bt}
Phillips J.~I., Wheeler C., Boylan-Kolchin M., Bullock J.~S., Cooper M.~C.,
  Tollerud E.~J., 2014{\natexlab{a}}, MNRAS, 437, 1930

\bibitem[{Phillips {et~al}\mbox{.}(2014{\natexlab{b}})Phillips, Wheeler,
  Cooper, Boylan-Kolchin, Bullock, \& Tollerud}]{Phillips:2014vt}
Phillips J.~I., Wheeler C., Cooper M.~C., Boylan-Kolchin M., Bullock J.~S.,
  Tollerud E., 2014{\natexlab{b}}, arXiv preprint, \eprint{1407.3276}

\bibitem[{Postman \& Geller(1984)}]{Postman:1984jx}
Postman M., Geller M.~J., 1984, ApJ, 281, 95

\bibitem[{Purcell {et~al}\mbox{.}(2007)Purcell, Bullock, \&
  Zentner}]{Purcell:2007cn}
Purcell C.~W., Bullock J.~S., Zentner A.~R., 2007, ApJ, 666, 20

\bibitem[{Ramella {et~al}\mbox{.}(1989)Ramella, Geller, \&
  Huchra}]{Ramella:1989fz}
Ramella M., Geller M.~J., Huchra J.~P., 1989, ApJ, 344, 57

\bibitem[{Ramella {et~al}\mbox{.}(2002)Ramella, Geller, Pisani, \&
  da~Costa}]{Ramella:2002id}
Ramella M., Geller M.~J., Pisani A., da~Costa L.~N., 2002, AJ, 123, 2976

\bibitem[{Ramella {et~al}\mbox{.}(1997)Ramella, Pisani, \&
  Geller}]{Ramella:1997kj}
Ramella M., Pisani A., Geller M.~J., 1997, AJ, 113, 483

\bibitem[{Ramella {et~al}\mbox{.}(1999)Ramella, Zamorani, Zucca, Stirpe,
  Vettolani, Balkowski, Blanchard, Cappi, Cayatte, Chincarini, Collins, Guzzo,
  MacGillivray, Maccagni, Maurogordato, Merighi, Mignoli, Pisani, Proust, \&
  Scaramella}]{Ramella:1999ul}
Ramella M. {et~al.}, 1999, A{\&}A, 342, 1

\bibitem[{Reddick {et~al}\mbox{.}(2013)Reddick, Wechsler, Tinker, \&
  Behroozi}]{Reddick:2013gi}
Reddick R.~M., Wechsler R.~H., Tinker J.~L., Behroozi P.~S., 2013, ApJ, 771, 30

\bibitem[{Robotham {et~al}\mbox{.}(2011)Robotham, Norberg, Driver, Baldry,
  Bamford, Hopkins, Liske, Loveday, Merson, Peacock, Brough, Cameron,
  Conselice, Croom, Frenk, Gunawardhana, Hill, Jones, Kelvin, Kuijken, Nichol,
  Parkinson, Pimbblet, Phillipps, Popescu, Prescott, Sharp, Sutherland, Taylor,
  Thomas, Tuffs, van Kampen, \& Wijesinghe}]{Robotham:2011hu}
Robotham A. S.~G. {et~al.}, 2011, MNRAS, 416, 2640

\bibitem[{Rodriguez-Puebla {et~al}\mbox{.}(2012)Rodriguez-Puebla, Drory, \&
  Avila-Reese}]{RodriguezPuebla:2012ku}
Rodriguez-Puebla A., Drory N., Avila-Reese V., 2012, ApJ, 756, 2

\bibitem[{Shankar {et~al}\mbox{.}(2006)Shankar, Lapi, Salucci, de~Zotti, \&
  Danese}]{Shankar:2006cd}
Shankar F., Lapi A., Salucci P., de~Zotti G., Danese L., 2006, ApJ, 643, 14

\bibitem[{Shen {et~al}\mbox{.}(2014)Shen, Yang, Mo, van~den Bosch, \&
  More}]{Shen:2014kq}
Shen S., Yang X., Mo H., van~den Bosch F., More S., 2014, ApJ, 782, 23

\bibitem[{Skibba(2009)}]{Skibba:2009kv}
Skibba R.~A., 2009, MNRAS, 392, 1467

\bibitem[{Skibba {et~al}\mbox{.}(2007)Skibba, Sheth, \&
  Martino}]{Skibba:2007fh}
Skibba R.~A., Sheth R.~K., Martino M.~C., 2007, MNRAS, 382, 1940

\bibitem[{Skibba {et~al}\mbox{.}(2011)Skibba, van~den Bosch, Yang, More, Mo, \&
  Fontanot}]{Skibba:2011dx}
Skibba R.~A., van~den Bosch F.~C., Yang X., More S., Mo H., Fontanot F., 2011,
  MNRAS, 410, 417

\bibitem[{Tago {et~al}\mbox{.}(2008)Tago, Einasto, Saar, Tempel, Einasto,
  Vennik, \& M{\"u}ller}]{Tago:2008kf}
Tago E., Einasto J., Saar E., Tempel E., Einasto M., Vennik J., M{\"u}ller V.,
  2008, A{\&}A, 479, 927

\bibitem[{Tago {et~al}\mbox{.}(2010)Tago, Saar, Tempel, Einasto, Einasto,
  Nurmi, \& Hein{\"a}m{\"a}ki}]{Tago:2010jd}
Tago E., Saar E., Tempel E., Einasto J., Einasto M., Nurmi P.,
  Hein{\"a}m{\"a}ki P., 2010, A{\&}A, 514, 102

\bibitem[{Tasitsiomi {et~al}\mbox{.}(2004)Tasitsiomi, Kravtsov, Wechsler, \&
  Primack}]{Tasitsiomi:2004gg}
Tasitsiomi A., Kravtsov A.~V., Wechsler R.~H., Primack J.~R., 2004, ApJ, 614,
  533

\bibitem[{Taylor {et~al}\mbox{.}(2015)Taylor, Hopkins, Baldry, Bland-Hawthorn,
  Brown, Colless, Driver, Norberg, Robotham, Alpaslan, Brough, Cluver,
  Gunawardhana, Kelvin, Liske, Conselice, Croom, Foster, Jarrett, Lara-Lopez,
  \& Loveday}]{Taylor:2015im}
Taylor E.~N. {et~al.}, 2015, MNRAS, 446, 2144

\bibitem[{Tempel {et~al}\mbox{.}(2012)Tempel, Tago, \&
  Liivam{\"a}gi}]{Tempel:2012er}
Tempel E., Tago E., Liivam{\"a}gi L.~J., 2012, A{\&}A, 540, 106

\bibitem[{Tempel {et~al}\mbox{.}(2014)Tempel, Tamm, Gramann, Tuvikene,
  Liivam{\"a}gi, Suhhonenko, Kipper, Einasto, \& Saar}]{Tempel:2014kg}
Tempel E. {et~al.}, 2014, A{\&}A, 566, 1

\bibitem[{Teyssier {et~al}\mbox{.}(2012)Teyssier, Johnston, \&
  Kuhlen}]{Teyssier:2012ky}
Teyssier M., Johnston K.~V., Kuhlen M., 2012, MNRAS, 426, 1808

\bibitem[{Tinker {et~al}\mbox{.}(2008)Tinker, Tinker, Kravtsov, Klypin, Klypin,
  Abazajian, Abazajian, Warren, Warren, Yepes, Yepes, Gottl{\"o}ber,
  Gottl{\"o}ber, Holz, \& Holz}]{Tinker:2008ja}
Tinker J. {et~al.}, 2008, ApJ, 688, 709

\bibitem[{Tinker {et~al}\mbox{.}(2011)Tinker, Wetzel, \&
  Conroy}]{Tinker:2011ek}
Tinker J., Wetzel A., Conroy C., 2011, \eprint{1107.5046}

\bibitem[{Tinker \& Wetzel(2010)}]{Tinker:2010ha}
Tinker J.~L., Wetzel A.~R., 2010, ApJ, 719, 88

\bibitem[{Trujillo-Gomez {et~al}\mbox{.}(2011)Trujillo-Gomez, Klypin, Primack,
  \& Romanowsky}]{TrujilloGomez:2011js}
Trujillo-Gomez S., Klypin A., Primack J., Romanowsky A.~J., 2011, ApJ, 742, 16

\bibitem[{Tucker {et~al}\mbox{.}(2000)Tucker, Oemler, Hashimoto, Shectman,
  Kirshner, Lin, Landy, Schechter, \& Allam}]{Tucker:2000ig}
Tucker D.~L. {et~al.}, 2000, ApJS, 130, 237

\bibitem[{Turner \& Gott(1976)}]{Turner:1976dw}
Turner E.~L., Gott J. R.~I., 1976, ApJS, 32, 409

\bibitem[{Vale \& Ostriker(2004)}]{Vale:2004bb}
Vale A., Ostriker J.~P., 2004, MNRAS, 353, 189

\bibitem[{van~den Bosch {et~al}\mbox{.}(2008)van~den Bosch, Aquino, Yang, Mo,
  Pasquali, McIntosh, Weinmann, \& Kang}]{vandenBosch:2008fv}
van~den Bosch F.~C., Aquino D., Yang X., Mo H.~J., Pasquali A., McIntosh D.~H.,
  Weinmann S.~M., Kang X., 2008, MNRAS, 387, 79

\bibitem[{van~den Bosch \& Jiang(2014)}]{vandenBosch:2014tl}
van~den Bosch F.~C., Jiang F., 2014, \eprint{1403.6835}

\bibitem[{van~den Bosch {et~al}\mbox{.}(2005)van~den Bosch, Weinmann, Yang, Mo,
  Li, \& Jing}]{vandenBosch:2005jo}
van~den Bosch F.~C., Weinmann S.~M., Yang X., Mo H.~J., Li C., Jing Y.~P.,
  2005, MNRAS, 361, 1203

\bibitem[{von~der Linden {et~al}\mbox{.}(2007)von~der Linden, Best, Kauffmann,
  \& White}]{vonderLinden:2007ev}
von~der Linden A., Best P.~N., Kauffmann G., White S. D.~M., 2007, MNRAS, 379,
  867

\bibitem[{Watson {et~al}\mbox{.}(2012)Watson, Berlind, McBride, Hogg, \&
  Jiang}]{Watson:2012jp}
Watson D.~F., Berlind A.~A., McBride C.~K., Hogg D.~W., Jiang T., 2012, ApJ,
  749, 83

\bibitem[{Watson {et~al}\mbox{.}(2015)Watson, Hearin, Berlind, Becker,
  Behroozi, Skibba, Reyes, Zentner, \& van~den Bosch}]{Watson:2015gq}
Watson D.~F. {et~al.}, 2015, MNRAS, 446, 651

\bibitem[{Weinmann {et~al}\mbox{.}(2009)Weinmann, Kauffmann, van~den Bosch,
  Pasquali, McIntosh, Mo, Yang, \& Guo}]{Weinmann:2009gf}
Weinmann S.~M., Kauffmann G., van~den Bosch F.~C., Pasquali A., McIntosh D.~H.,
  Mo H., Yang X., Guo Y., 2009, MNRAS, 394, 1213

\bibitem[{Weinmann {et~al}\mbox{.}(2006)Weinmann, van~den Bosch, Yang, \&
  Mo}]{Weinmann:2006hu}
Weinmann S.~M., van~den Bosch F.~C., Yang X., Mo H.~J., 2006, MNRAS, 366, 2

\bibitem[{Wetzel {et~al}\mbox{.}(2014)Wetzel, Tinker, Conroy, \&
  Bosch}]{Wetzel:2014up}
Wetzel A.~R., Tinker J.~L., Conroy C., Bosch F. C. v.~d., 2014, MNRAS, 439,
  2687

\bibitem[{Wetzel {et~al}\mbox{.}(2013)Wetzel, Tinker, Conroy, \& van~den
  Bosch}]{Wetzel:2013dw}
Wetzel A.~R., Tinker J.~L., Conroy C., van~den Bosch F.~C., 2013, MNRAS, 432,
  336

\bibitem[{Wetzel {et~al}\mbox{.}(2012)Wetzel, Wetzel, Tinker, Tinker, Conroy,
  \& Conroy}]{Wetzel:2012lk}
Wetzel A.~R., Wetzel A.~R., Tinker J.~L., Tinker J.~L., Conroy C., Conroy C.,
  2012, MNRAS, 424, 232

\bibitem[{Yang {et~al}\mbox{.}(2008)Yang, Mo, \& van~den Bosch}]{Yang:2008tg}
Yang X., Mo H.~J., van~den Bosch F.~C., 2008, ApJ, 676, 248

\bibitem[{Yang {et~al}\mbox{.}(2009)Yang, Mo, \& van~den Bosch}]{Yang:2009cm}
Yang X., Mo H.~J., van~den Bosch F.~C., 2009, ApJ, 695, 900

\bibitem[{Yang {et~al}\mbox{.}(2005)Yang, Mo, van~den Bosch, \&
  Jing}]{Yang:2005cn}
Yang X., Mo H.~J., van~den Bosch F.~C., Jing Y.~P., 2005, MNRAS, 356, 1293

\bibitem[{Yang {et~al}\mbox{.}(2007)Yang, Mo, van~den Bosch, Pasquali, Li, \&
  Barden}]{Yang:2007db}
Yang X., Mo H.~J., van~den Bosch F.~C., Pasquali A., Li C., Barden M., 2007,
  ApJ, 671, 153

\bibitem[{York {et~al}\mbox{.}(2000)York, Adelman, Anderson, Anderson, Annis,
  Bahcall, Bakken, Barkhouser, Bastian, Berman, Boroski, Bracker, Briegel,
  Briggs, Brinkmann, Brunner, Burles, Carey, Carr, Castander, Chen, Colestock,
  Connolly, Crocker, Csabai, Czarapata, Davis, Doi, Dombeck, Eisenstein,
  Ellman, Elms, Evans, Fan, Federwitz, Fiscelli, Friedman, Frieman, Fukugita,
  Gillespie, Gunn, Gurbani, de~Haas, Haldeman, Harris, Hayes, Heckman,
  Hennessy, Hindsley, Holm, Holmgren, Huang, Hull, Husby, Ichikawa, Ichikawa,
  Ivezi{\'c}, Kent, Kim, Kinney, Klaene, Kleinman, Kleinman, Knapp, Korienek,
  Kron, Kunszt, Lamb, Lee, Leger, Limmongkol, Lindenmeyer, Long, Loomis,
  Loveday, Lucinio, Lupton, MacKinnon, Mannery, Mantsch, Margon, McGehee,
  McKay, Meiksin, Merelli, Monet, Munn, Narayanan, Nash, Neilsen, Neswold,
  Newberg, Nichol, Nicinski, Nonino, Okada, Okamura, Ostriker, Owen, Pauls,
  Peoples, Peterson, Petravick, Pier, Pope, Pordes, Prosapio, Rechenmacher,
  Quinn, Richards, Richmond, Rivetta, Rockosi, Ruthmansdorfer, Sandford,
  Schlegel, Schneider, Sekiguchi, Sergey, Shimasaku, Siegmund, Smee, Smith,
  Snedden, Stone, Stoughton, Strauss, Stubbs, SubbaRao, Szalay, Szapudi,
  Szokoly, Thakar, Tremonti, Tucker, Uomoto, Vanden~Berk, Vogeley, Waddell,
  Wang, Watanabe, Weinberg, Yanny, Yasuda, \& Collaboration}]{York:2000gn}
York D.~G. {et~al.}, 2000, AJ, 120, 1579

\bibitem[{Zabludoff \& Mulchaey(1998)}]{Zabludoff:1998gj}
Zabludoff A.~I., Mulchaey J.~S., 1998, ApJ, 496, 39

\bibitem[{Zehavi {et~al}\mbox{.}(2002)Zehavi, Blanton, Frieman, Weinberg, Mo,
  Strauss, Anderson, Annis, Bahcall, Bernardi, Briggs, Brinkmann, Burles,
  Carey, Castander, Connolly, Csabai, Dalcanton, Dodelson, Doi, Eisenstein,
  Evans, Finkbeiner, Friedman, Fukugita, Gunn, Hennessy, Hindsley, Ivezi{\'c},
  Kent, Knapp, Kron, Kunszt, Lamb, Leger, Long, Loveday, Lupton, McKay,
  Meiksin, Merrelli, Munn, Narayanan, Newcomb, Nichol, Owen, Peoples, Pope,
  Rockosi, Schlegel, Schneider, Scoccimarro, Sheth, Siegmund, Smee, Snir,
  Stebbins, Stoughton, SubbaRao, Szalay, Szapudi, Tegmark, Tucker, Uomoto,
  Vanden~Berk, Vogeley, Waddell, Yanny, \& York}]{Zehavi:2002dd}
Zehavi I. {et~al.}, 2002, ApJ, 571, 172

\bibitem[{Zentner {et~al}\mbox{.}(2014)Zentner, Hearin, \& van~den
  Bosch}]{Zentner:2014ki}
Zentner A.~R., Hearin A.~P., van~den Bosch F.~C., 2014, MNRAS, 443, 3044

\end{thebibliography}

\appendix

\section{Shuffled Mock}
\label{appendix:shuffling}

The age-matching based mock has galactic conformity build into it. As such, it is ideally suited to test whether galaxy group finders can recover this conformity signal, both qualitatively and quantitatively. However, we also would like to perform a null-test, based on a mock catalogue without galactic conformity. To do so, we construct a second mock derived from the age-matching mock by shuffling galaxy populations among haloes of of similar mass. 

In detail, the shuffling consists of the following steps. First we look up all haloes from the ROCKSTAR Bolshoi $z=0$ halo catalogue which did not receive a galaxy when applying the abundance matching technique described above and add these to our mock catalogue with null values for all galaxy properties. Next we bin haloes in small, $0.1$ dex, bins in halo mass.  We then shuffle entire galaxy groups, i.e. galaxies which occupy the same host halo, amongst haloes in a bin (allowing only one group per halo), preserving the relative positions between a central galaxy and its satellite(s)--note that this destroys the galaxy-sub-halo connection.  This shuffling removes any 2-halo correlation between galaxy groups (e.g., 2-halo conformity) present in the age-matching mock.  Next we remove all empty haloes from the new catalogue and apply one additional shuffling.  Haloes are again binned in small, $0.1$ dex bins in halo mass, and this time only satellite galaxies are shuffled among the haloes in a bin, whereby we preserve for each satellite galaxy its relative distance from the halo centre such that $\vec{x}_{\rm sat}-\vec{x}_{\rm cen}$ is constant throughout the shuffling process. This shuffling destroys any correlation between the properties of centrals and satellites (e.g., 1-halo conformity). In the text we refer to this mock as the ``shuffled mock'' and the unaltered version as the ``age-matching mock''.

The age-matching mock and shuffled mock have the exact same HOD, $P(N_{\rm gal}|M_{\rm vir})$, by design.  However, the clustering statistics of the galaxies change.  The auto 2-point correlation of galaxies in the age-matching mock is slightly higher than in the shuffled mock.  This is because we shuffle galaxies at fixed halo mass, and not fixed $V_{\rm peak}$, the value used to populate galaxies in haloes when building the age-matching mock \citep*[see][]{Zentner:2014ki}.

\section{Group Finding Errors: individual haloes and groups}
\label{appendix:group_finding_errors}

Throughout this section, we define the subscript `$\rm g$' to mean ``group member'' and `$\rm ng$' to mean ``not a group member''.  Similarly, we define the subscript `$\rm h$' to mean ``halo member'' and `$\rm nh$' to mean ``not a halo member''.  If one desires to compare the membership of a group to a halo, it is necessary to define a unique connection between every group and halo.  This is a non-trivial choice, and the decision may strongly influence the resulting group/halo level statistics of interest.  We do not make a specific choice for this discussion, but we will briefly discuss possible choices used in the literature.  

\begin{figure}
    \includegraphics[]{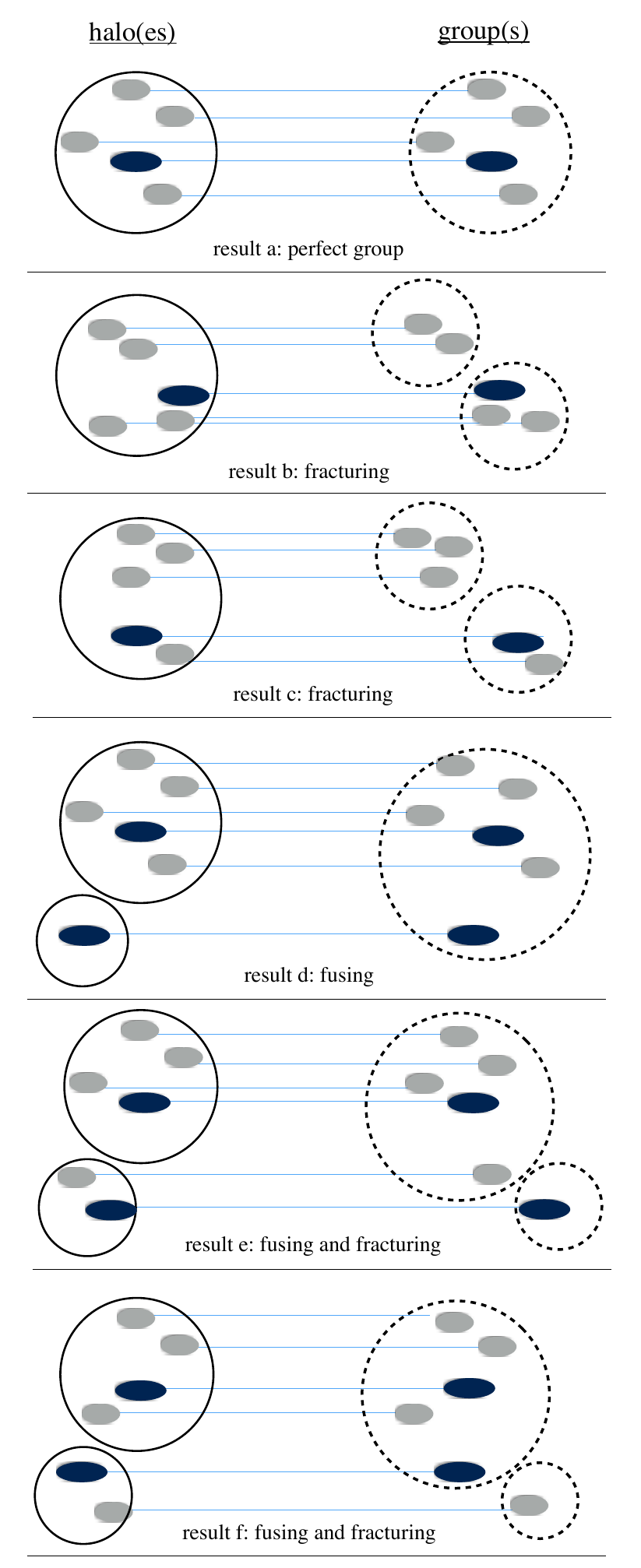} 
    \caption{diagram of the possible outcomes of placing galaxies which live in haloes(solid circles) into groups(dashed circles).  One goal of a group finder is to assign all galaxies which occupy a common halo, and only those galaxies, to a unique group(result a).  Two processes result in errors in the group finding process: fracturing--galaxies which live in a common halo assigned to multiple groups(result b,c), and fusing--galaxies from two distinct haloes assigned to a common group(result d).  These two processes may also occur simultaneously(result e,f).}
\label{fig:failure_diagram_1}
\end{figure}

One may define the halo associated with a group as the halo associated with a {\em true} central group member \citep[e.g.][]{Yang:2007db, MunozCuartas:2012cx}, where if there are more than one, the true central of the most massive halo (or as identified as a central in the group).  This will result in any group that does not contain a true central, to not have an associated halo (fig. \ref{fig:failure_diagram_1} result b,e,f).  Additionally, any group that contains more than one true central will cause one or more haloes to have no associated group (fig. \ref{fig:failure_diagram_1} result d,f).  Variants on this method are possible.  If more than one halo's central is in a group, \cite{MunozCuartas:2012cx} connect the group to the halo with the most members in the group.

Another possible choice is to define the group associated with a halo as the group associated with the largest number of halo members, or majority \citep[e.g.][]{Eke:2004bf, Robotham:2011hu}.  This definition may result in one or more groups being associated with no halo.  For example, any halo that has undergone a pure fracturing process during group finding, will produce two or more groups, where, for each group, all members belong to a common halo (fig. \ref{fig:failure_diagram_1} result b,c).  This may also happen for a combination of fusing and fracturing of haloes.  Additionally, a contingency should be defined for when two or more haloes have equal numbers of associated group members with no halo having more. 

Other options are of course possible.  \cite{Merchan:2002gp} associate groups to haloes by matching the centre of mass between groups in redshift space and haloes in real space, while \cite{Calvi:2011hp} match two group catalogues as a cross-check in a similar way.

Once a halo-group connection has been made, we can define the total number of group members as: 
\begin{equation}
N_{\rm g} = N_{\rm g|h} + N_{\rm g|nh}
\end{equation}
and the total number of halo members as:
\begin{equation}
N_{\rm h} = N_{\rm g|h} + N_{\rm ng|h}.
\end{equation}
We define the completeness of an individual group's membership as:
\begin{equation}
C_{\rm mem} = \frac{N_{\rm g|h}}{N_{\rm g|h}+N_{\rm ng|h}}= \frac{N_{\rm g|h}}{N_{\rm h}}
\end{equation}
and the purity as:
\begin{equation}
P_{\rm mem} = \frac{N_{\rm g|h}}{N_{\rm g|h}+N_{\rm g|nh}}= \frac{N_{\rm g|h}}{N_{\rm g}}.
\end{equation}
One more useful quantity is the transition factor:
\begin{equation}
T_{\rm mem} = \frac{N_{\rm g|h}+N_{\rm g|nh}}{N_{\rm g|h}+N_{\rm ng|h}} = \frac{C_{\rm mem}}{P_{\rm mem}}.
\end{equation}
Note that $(T_{\rm mem} = 1) \implies (C_{\rm mem}=P_{\rm mem})$. In the limit where $\langle T_{\rm mem}\rangle(M_h) = 1$ , the HOD may be recovered.  However, $T_{\rm mem}=1$ only implies purity equal completeness, it does not require the purity and completeness be near 1.  So, in general, it is desirable to tune a group finder such that $\left[ C_{\rm mem} \rightarrow 1, P_{\rm mem} \rightarrow 1, T_{\rm mem} \rightarrow 1 \right]$.  The ideal balance of these three conditions will depend on the goal of the study for which the group finder is being tuned. 

We choose not to use the approach outlined above as the choice of how to define a halo-group connection is arbitrary, and we expect that the choice will significantly affect the results of any purity and completeness analysis.

\section{Difficulties in defining groups in simulations}
\label{appendix:pernicious_effects}

\begin{figure}
    \includegraphics[]{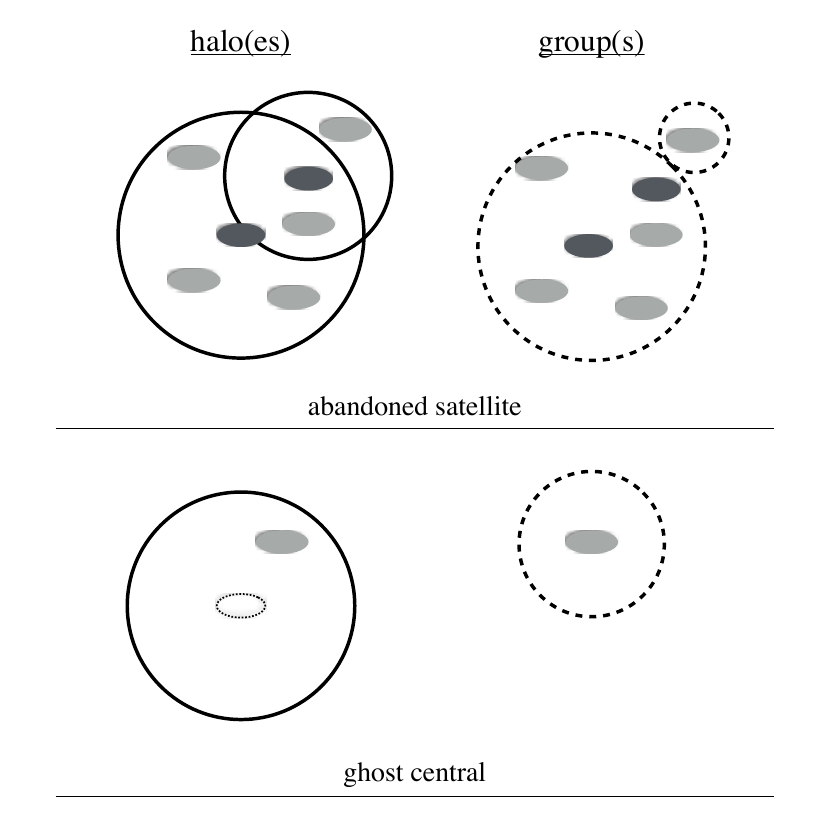} 
    \caption{diagram of the possible pernicious effects when placing galaxies which live in haloes(solid circles) into groups(dashed circles).  top: We show an abandoned satellite--If two haloes overlap, such that the central galaxy of the less massive halo falls within the virial volume of the larger halo, while one of its satellites does not, the satellite will occupy a group with no central, as its previous central is now a satellite of the larger halo.  bottom: We show a group with a ghost central--Because there is scatter in the $V_{\rm peak}-L_{\rm gal}$ relation, a halo's central galaxy may have a luminosity below the magnitude limit of the sample, while the satellite has a luminosity above the limit, resulting in a group with no central.}
\label{fig:failure_diagram_2}
\end{figure}

There are two phenomena present in sub-halo abundance matching (SHAM) mocks which one should be aware of when comparing the results of a group finder to a SHAM mock.  First, spherical over-density based halo finders run on N-body simulations to build the mock will sometimes disjoin a halo while it is being accreted into a new host halo (or more generally, while passing through the virial volume of a larger halo).  When this occurs, some particles of the now disjoint halo remain outside the virial volume of the new host halo.  This can have the effect of separating central and satellite populations that are part of one halo (or have been in the past), but which have been separated at a particular time-step in a halo catalogue.  A particularly pernicious outcome can be a group in the mock with no true central galaxy, because the central galaxy has been accreted into a new host halo, while one or more of its satellites have not (see top panel of Fig. \ref{fig:failure_diagram_2}).  We refer to these satellites as abandoned satellites.  When a group exists that consists entirely of abandoned satellites, it violates the assumption that every group must have a central galaxy.  This creates a problem when comparing the result of a group finder to the mock, as the mock has already violated an ansatz of the group finder, namely that every group must have a central galaxy.

There is a second effect to be aware of when comparing a mock to group finder results which occurs when defining a volume limited complete galaxy sample.  It is possible to create groups with a satellite and no central when a central galaxy of a group does not make the cut, while a satellite does.  We refer to this phenomenon as a ``ghost'' central.  This results in a group with no central, but only in the sense that no central made the sample cut.  Never-the-less, such a  group also violates the assumption that every group must have a central galaxy.  

We do not account for either of these effects when calculating group statistics or when shuffling the mock to create our shuffled mock catalogue (described in \ref{appendix:shuffling}).  These two effects together affect $\sim 1\%$ of all satellites in the age-matching mock, with there being approximately equal numbers of abandoned satellites and satellites with a ghost central.

\bsp

\label{lastpage}

\end{document}